\pdfoutput=1
\documentclass[12pt, a4paper]{article}
\usepackage{graphicx}
\usepackage{amsmath,amssymb}

\newcommand{\be}{\begin{equation}}
\newcommand{\ee}{\end{equation}}
\newcommand{\ba}{\begin{eqnarray}}
\newcommand{\ea}{\end{eqnarray}}

\renewcommand{\l}{\left(}
\renewcommand{\r}{\right)}

\newcommand{\half}{\frac{1}{2}}

\begin{document}
\rightline{INR-TH/2015-016}

\begin{center}

{\bf \LARGE Nucleon-decay-like signatures of Hylogenesis} \\
\bigskip
{\large
S.\,V.\,Demidov$^{a}$\footnote{{\bf e-mail}: demidov@ms2.inr.ac.ru},
D.\,S.\,Gorbunov$^{a,b}$\footnote{{\bf e-mail}: gorby@ms2.inr.ac.ru}\\
\vspace{10pt}
$^a$\textit{Institute for Nuclear Research of the Russian Academy
of Sciences, Moscow 117312, Russia}\\
\vspace{5pt}
$^b$\textit{Moscow Institute of Physics and Technology, 
  Dolgoprudny 141700, Russia}%
}
\end{center}

\begin{abstract}
We consider nucleon-decay-like signatures of the hylogenesis, a
variant of the antibaryonic dark matter model.  For the interaction
between visible and dark matter sectors through the neutron portal, we
calculate the rates of dark matter scatterings off neutron which mimic
neutron-decay processes $n\to \nu\gamma$ and $n\to \nu e^+ e^-$ with
richer kinematics. We obtain bounds on the model parameters from
nonobservation of the neutron decays by applying the kinematical cuts
adopted in the experimental analyses. The bounds are generally 
(much) weaker than those coming from the recently performed
study of events with a single jet of high
transverse momentum and missing energy observed at the LHC.  Then we
suggest several new nucleon-decay like processes with two mesons in
the final state and estimate (accounting for the LHC constraints)  
the lower limits on the nucleon lifetime with respect to these
channels. The obtained values appear to be promising for probing the
antibaryonic dark matter at future underground experiments like HyperK
and DUNE.
\end{abstract}

\newpage
\section{Introduction}
\label{Sec:Intro}

Given a variety of spatial scales and cosmological epochs associated
with dark matter phenomena, their natural explanation seems in
introducing a new neutral particle, stable at cosmological
time-scales. Many extensions of the Standard Model of particle physics 
(SM) suggest suitable dark matter candidates with masses ranging from
$10^{-23}$\,eV (oscillating scalar field, see e.g.\,\cite{Hu:2000ke})
to $10^{16}$\,GeV 
(superheavy dark matter, see e.g.\,\cite{Gorbunov:2012ij}).  Dark
matter particles must be produced in the early Universe at a stage
before matter-radiation equality. Most mechanisms exploited for
this purpose work properly (for a review see\,\cite{Baer:2014eja}) but
treat the (order-of-magnitude) equality of dark matter and
visible matter contributions to the present energy density of
Universe,  
\begin{equation}
\label{parity}
\rho_{\text{DM},0}\sim\rho_{\text{B},0}\,,
\end{equation}
as an accidental coincidence.

Yet it may be a hint towards a common origin of both cosmological
problems, dark matter phenomena and matter-antimatter asymmetry of the
Universe. There are models addressing this issue. In particular, an elegant approach
is provided by models of antibaryonic dark matter, where dark matter
particles carry (anti)baryonic charge. 
The idea is that the total baryonic charge of
the Universe is zero, but it is redistributed between visible sector
(positive baryonic charge) and dark sector (negative charge of the
same amount). Both dark and visible matter emerge during the same
process at some stage in the early Universe making a connection 
between the two components, so that the coincidence~\eqref{parity} may
be understood.   

Similar to the visible sector, the dark sector is asymmetric, being
populated solely with particles of negative baryonic charge. The
models of this type are called asymmetric dark matter, for 
a review see \cite{Petraki:2013wwa}. They exhibit quite specific
phenomenology. As a rule, no dark matter pair annihilation is expected
in galaxies or inside the Sun (see,
however,~\cite{Hardy:2014dea,Bell:2014xta}).  
Instead, the antibaryonic dark matter particle may annihilate with
nucleon, mimicking proton/neutron disappearance or decay. 

A remarkable example of the antibaryonic dark matter model is
provided by hylogenesis\,\cite{Davoudiasl:2010am}. To the SM particle
content at low energies the model adds a complex scalar $\Phi$ and Dirac 
spinor $\Psi$, together forming dark matter components, and also two
heavy fermions 
$X_a$, $a=1,2$, playing the role of messengers between the visible and
dark sectors. The interaction terms read 
\begin{equation}
\label{main-lagr}
{\cal L}=-\frac{\lambda_a^{ijk}}{\Lambda^2}\, \overline{X_a}
\,\frac{1+\gamma_5}{2}\,d^i\cdot \overline{u^{j\,C}}\,\frac{1+\gamma_5}{2}\, 
d^k+\zeta_a\overline{X_a} \Psi^C\Phi^* + \text{h.c.}\,,  
\end{equation}
with $i,j,k$ running over the SM three generations, $d^i$ and $u^j$
denote down-type and up-type quarks, superscript $C$ refers to charge
conjugation; $\lambda_a^{ijk}$ and $\zeta_a$ are dimensionless
coupling constants, $\Lambda$ stands for the scale of new physics
which completes the model to a renormalizable theory (for a particular
variant of high-energy completion within a supersymmetric 
framework see\,\cite{Blinov:2012hq}).  

The new fields carry baryonic charge, so that $B(X_a)=1$ and
$B(\Psi)=B(\Phi)=-1/2$. Coupling constants $\lambda_a^{ijk}$ and
$\zeta_a$ are, in general, complex numbers, providing the model with
charge ($C$) and charge-parity ($CP$) violation required for
the successful dynamical generation 
of the baryon asymmetry. The latter is produced in the early Universe
via $CP$-violating decays of nonrelativistic messengers $X_a$ in a way 
very similar to what happens in the standard leptogenesis with heavy
sterile neutrinos \cite{Fukugita:1986hr}. Since the baryon number is
conserved by interactions~\eqref{main-lagr}, in the same process the
dark sector ($\Psi$, $\Phi$) becomes asymmetric, collecting the
negative baryonic charge produced in the $CP$-violating decays of
nonrelativistic fermions $X_a$. Later in the Universe, baryons and
antibaryons of the visible sector 
annihilate, leaving the net baryonic charge, which is accumulated at
present mostly in hydrogen and helium. A similar process happens in
the dark 
sector, and the antibaryonic charge of the same amount is distributed
between fermions $\Psi$ and bosons $\Phi$. This may be characterized
by a ratio of their present number densities, 
\begin{equation}
\label{ratio}
\eta\equiv \frac{n_{\Phi\,,0}}{n_{\Psi\,,0}}\,.
\end{equation}

Proton and both dark matter particles, $\Psi$ and $\Phi$, are stable,
if their masses obey the kinematical constraints
\begin{equation}
\label{kinematic-constraint}
\left| M_\Psi-M_\Phi\right|<M_p+m_e < M_\Psi + M_\Phi\,,
\end{equation}
where $M_p$ and $m_e$ stand for proton and electron masses. Total
baryon number conservation implies a simple relation between dark 
matter and visible baryon number densities 
\begin{equation}
\label{number-density}
n_B=\frac{n_\Psi+n_\Phi}{2}\,.
\end{equation}
For the present dark matter energy density, one can write 
\begin{equation}
\label{energy-density}
\rho_{DM,0}=M_\Psi\, n_\Psi+ M_\Phi\,n_\Phi\,.
\end{equation}
Without any asymmetry between the two dark matter components,
i.e., when $\eta=1$, we obtain from~\eqref{energy-density}
and~\eqref{number-density}  
\begin{equation}
\label{relation-symmetric}
\rho_{DM,0}=\frac{M_\Psi+M_\Phi}{M_p}\,\rho_{B,0}\,,
\end{equation}
which for the observed property~\eqref{parity} settles the dark matter 
mass scale in the GeV-range. Then, for the present cosmological
estimates 
of $\rho_{B,0}$ and $\rho_{DM,0}$ \cite{Agashe:2014kda}, the sum of the
dark matter particle masses are fixed by
eq.\,\eqref{relation-symmetric}, while the kinematical
constraint\,\eqref{kinematic-constraint} confines the individual
masses inside the interval 
\begin{equation}
\label{interval-symmetric}
1.7\,\text{GeV}\lesssim M_\Psi\,, \, M_\Phi \lesssim 2.9\,\text{GeV}\,. 
\end{equation} 
With asymmetry between $\Psi$ and $\Phi$ populations, $\eta\neq 1$, the
relation\,\eqref{relation-symmetric} is replaced with 
\begin{equation}
\label{relation-asymmetric}
\rho_{DM,0}=\frac{2(M_\Psi+\eta\,M_\Phi)}{(1+\eta)M_p}\,\rho_{B,0}\,. 
\end{equation}

The interaction with quarks in \eqref{main-lagr} can be used to probe
the model at
colliders\,\cite{Davoudiasl:2011fj,Demidov:2014mda}. Heavy fermions
$X_a$ can be directly produced or virtually contribute to dark matter
production. This model provides the following signatures for the LHC
experiments (depending on the quark structure in\,\eqref{main-lagr}):
{\it (i)} missing energy and either a jet with high 
transverse momentum $p_T$\,\cite{Davoudiasl:2011fj,Demidov:2014mda} or
a heavy quark ($t,b,$ or $c$) with high $p_T$\,\cite{Demidov:2014mda};
{\it (ii)} a jet (or a heavy quark) with high $p_T$ and a peak in the 
invariant mass of three jets whose momenta compensate high
$p_T$\,\cite{Demidov:2014mda}. The performed analysis of LHC events
with a high-$p_T$ jet and missing energy has allowed us to constrain
the model parameter space pushing the new physics up to TeV
scale\,\cite{Demidov:2014mda}.

Another very pronounced signature of the
model \,\cite{Davoudiasl:2010am} is an induced nucleon 
decay (IND)~\cite{Davoudiasl:2011fj}. The dark matter particle
scattering  off a nucleon (through the 
exchange of virtual fermions $X_a$) flips its type,
$\Psi\leftrightarrow\Phi$, and destroys the nucleon. The kinematical
constraint\,\eqref{kinematic-constraint} obviously forbids the
traceless disappearance of the nucleon, i.e., a process like
$\Phi+n\to\Psi$. Some additional particles must emerge in the final
state 
yielding a signature of the induced nucleon decay. These processes
involving an additional single meson in the final state have been
analyzed 
\cite{Davoudiasl:2010am,Davoudiasl:2011fj,Blinov:2012hq} 
for a set of quark operators entering
\eqref{main-lagr} and a number of final states. While the scattering
mimics the nucleon decay, the kinematics of particles in the final
state is different, which prevents us from direct use of the limits on 
the proton/neutron lifetimes to constrain the model parameter
space. However, by adjusting properly the kinematical cuts, the
corresponding analysis has been
performed\,\cite{Blinov:2012hq,Davoudiasl:2011fj}. In particular, for
$X_a$ couplings to the $uds$ operator in \eqref{main-lagr}, the
results of nucleon decay searches raise the mass of heavy fermion
$X_a$ and the scale of new physics $\Lambda$ up to the TeV
scale\,\cite{Davoudiasl:2010am,Davoudiasl:2011fj,Blinov:2012hq}.  

In this paper we analyze several new modes of the induced nucleon
decays via a neutron 
portal, represented by $dud$ operator in~\eqref{main-lagr}. The paper 
is organized as follows. In Sec.\,\ref{Sec:Lagr} we derive the low
energy effective lagrangian describing the dark matter scattering off
a neutron and give the relation between the scattering  cross section
and the nucleon lifetime with respect to decay into a given final
state. In Sec.\,\ref{Sec:gamma} we consider $2\to 2$ scattering
processes $\Psi(\Phi) + n\to \Phi(\Psi) + \gamma$, which mimic neutron
decay $n\to \nu\gamma$, and, imposing the cuts adopted in the
experimental search for this decay
mode \cite{Blewitt:1985zg,McGrew:1999nd}, we constrain the model
parameter space. These constraints turn out to be 
(much) weaker  than those following from the
LHC\,\cite{Demidov:2014mda}, 
so finally we obtain a lower estimate of the neutron lifetime in this
model based on the limits from the LHC. In a similar way, we
investigate the scattering $\Psi(\Phi) + n\to \Phi(\Psi) + e^+\, e^-$
in Sec.\,\ref{Sec:ee}. We study the induced nucleon decays into two
light mesons ($\pi$, $K$, $\eta$ in various possible combinations) in
Sec.\,\ref{Sec:2mesons} and (based 
on the LHC bounds\,\cite{Demidov:2014mda}) predict the shortest
lifetimes at the level of $10^{32}-10^{33}$\,yr expected for
these modes within hylogenesis. The obtained numbers are quite
promising and allow the processes to be tested with the next
generation underground facilities like
HyperK\,\cite{HyperK,Abe:2011ts} and DUNE\,\cite{DUNE}. 
We expect that these channels apart from the dominant
single-meson-induced nucleon decays would be helpful to discriminate
between different models predicting processes with baryon number
violation and corner an interesting region in the parameter space of
the hylogenesis scenario if a signal of nucleon-decay-type is
found in future. We conclude in Sec.\,\ref{Sec:conclusion}.

\section{Low-energy effective lagrangian and nucleon lifetime}
\label{Sec:Lagr}

The coupling terms in eq.\,\eqref{main-lagr} relevant for low-energy
phenomenology of the neutron portal read 
\begin{equation}
\label{eff-lagr}
{\cal L}=-\frac{\lambda_a^{dud}}{\Lambda^2}\, \overline{X_a}
\,\frac{1+\gamma_5}{2}\,d\cdot \overline{u^C}\,\frac{1+\gamma_5}{2}\, 
d+\zeta_a\overline{X_a} \Psi^C\Phi^* + \text{h.c.}\,.  
\end{equation}
Hereafter we are interested in processes with typical energies much
below the mass scale of the heavy fermions $X_a$. The exchange of
virtual $X_a$ between the visible sector and dark sector fields
entering \eqref{eff-lagr} yields the following contact interaction 
\begin{equation}
\label{contact}
{\cal L}=-\frac{\sum_{a=1}^2\frac{\lambda_a^{dud}\,\zeta_a^*}{M_{X_a}}}{\Lambda^2}\,
\Phi\,\overline{\Psi^C} \frac{1+\gamma_5}{2}\,d\cdot \overline{u^C}\,\frac{1+\gamma_5}{2}\, 
d + \text{h.c.}\,.  
\end{equation}
For further analysis it is convenient to introduce variables
$M_X$ and $y$   by relations
\begin{equation}
\label{new-variables}
\frac{y}{M_X}\equiv
\sum_{a=1}^2\frac{\lambda_a^{dud}\,\zeta_a^*}{M_{X_a}}\,, 
\end{equation}
so that $M_X$ (somewhat vaguely) indicates the heavy fermion
scale, while dimensionless parameter $y$ reflects the coupling
strength. The physical meaning of $M_X$ is the energy scale below
which the effective interaction \eqref{contact} can be safely
exploited instead of \eqref{eff-lagr}. Since not $y$ and $M_X$
individually  but only their ratio~\eqref{new-variables} 
enters all the formulas below, there is an ambiguity in the
definition of $y$ and $M_X$ related to the change of the variables. 
However, it has no impact on the physical observables. 

Further, the GeV scale of dark matter masses
\eqref{interval-symmetric} and smallness of the expected velocity of
galactic dark matter particles allow us to describe the dark matter
scattering off nucleons in terms of baryons and mesons rather than
quarks and gluons. In this approximation, the lagrangian
\eqref{contact} with replacement \eqref{new-variables} transforms
into Yukawa-type interaction 
\begin{equation}
\label{Yukawa}
{\cal L}=-\frac{y\,\beta}{\Lambda^2\,M_X}\,
\Phi\,\overline{\Psi^C} \frac{1+\gamma_5}{2}\,n + \text{h.c.}\,,
\end{equation}
which we use below to calculate the scattering rates; $n$ denotes
the neutron field and the parameter $\beta=0.012$\,GeV$^3$ is related
to the QCD scale\,\cite{Claudson:1981gh}.   

The cross sections of dark matter scatterings off nucleon $N$,
$\sigma_{\Psi N\to\dots}$ and $\sigma_{\Phi N\to \dots}$, are related
to the total nucleon lifetime with respect to a particular IND process
$\tau_{N\to\dots}$ as follows
\begin{equation}
\label{lifetime}
\tau_{N\to\dots}=\frac{1}{n_\Psi\, v \sigma_{\Psi N\to\dots} +n_\Phi\, v \sigma_{\Phi N\to \dots}}\,,
\end{equation}
where $v$ is the dark matter particle velocity in the laboratory
frame  where nucleons are at rest. In fact, since the scatterings we
discuss happen in $s$-wave, the cross sections are inversely
proportional to $v$, and the lifetime \eqref{lifetime} does not depend
on its value.

\section{Scattering processes $\Psi(\Phi)\,
n\to \Phi(\Psi)\, \gamma$} 
\label{Sec:gamma}

We start our study with a simple $2\to 2$ scattering with dark matter
particles annihilating a neutron into dark matter particle of another
type and a photon. Let $p_\Psi$, $p_n$ and $q$ be the 4-momentum of
$\Psi$, neutron $n$, and the outgoing photon $\gamma$, being real for
$\Psi\, n\to \Phi\, \gamma$, and hence $q^2=0$ (or virtual for $\Psi\, 
n\to \Phi\, e^+e^-$, which we consider in Sec.\,\ref{Sec:ee}). 
The process is proceeded due to the Yukawa interaction
\eqref{Yukawa} and the neutron dipole moment 
\begin{equation}
\label{lagr2}
{\cal L}= \frac{ie}{2M_n} \, \bar n \sigma^{\mu\nu}q_\nu F_2(q^2) n A_\mu\;,
\end{equation}
where $\epsilon_\mu(q)$ is photon polarization
4-vector and for the Pauli (magnetic) form factor
we utilize the dipole parametrization 
$F_2(q^2)=-1.91/(1+q^2r_M^2/12)^2$ with
magnetic radius $r_M=0.86$\,fm \cite{Agashe:2014kda}.  

The dark matter particle scatters off the neutron by means of virtual
neutron exchange.  The matrix element of the process reads 
\[
\frac{iey\beta }{2M_n\,\Lambda^2}\frac{F_2(q^2)}{M_X} 
\, \overline{\Psi^C}(p_\Psi)\frac{1+\gamma_5}{2}\frac{\hat p
-M_n}{p^2-M_n^2} \sigma^{\mu\nu}q_\nu n (p_n) \epsilon_\mu(q)\Phi,
\]
where $p=q-p_n$ is the 4-momentum of the virtual neutron, and 
$n(p_n), \Psi(p_\Psi), \Phi$ are wave functions of the neutron, $\Psi$
and $\Phi$ particles, respectively. In the laboratory frame the
neutron is at rest, while the dark matter particle moves with small
velocity $v\ll~1$. Here and below, we perform the estimates to the
leading order in velocity $v$. The squared matrix element averaged
over spins of the two incoming fermions in the laboratory frame is  
\begin{equation}
\label{n-gamma}
\overline{\left| {\cal M}\right|^2}= 
\frac{e^2\, y^2\, \beta^2}{
4M_n^2}\frac{F^2_2(0)}{M_X^2\,\Lambda^4} 
\, M_n\, M_\Psi \,.
\end{equation}

For the similar process $\Phi\, n \to \Psi\, \gamma$, we find the same
expression \eqref{n-gamma} up to the following replacement 
\[
M_n\,M_\Psi \to 2M_n\, (M_\Phi + M_n - q_0),
\]
where the additional factor accounts for different numbers of fermions
in the initial states averaged over spins. To the leading order in
$v\ll 1$, the photon frequency is 
\[
q_0\approx M_n+M_\Psi-M_\Phi\;.
\]

We can place a bound on the model parameter space from
nonobservation of the decay $n\to \nu\gamma$
\cite{Blewitt:1985zg,McGrew:1999nd} exhibiting the same
signature as the scattering process under discussion: a single photon
in the final state. To this end, we constrain the kinematics of the
photon as it has been  adopted\,\footnote{One more requirement on the
quantity called {\it asymmetry} to be discussed in Sec.\,\ref{Sec:ee}
is automatically fulfilled.} 
in the original 
experimental analysis \cite{Blewitt:1985zg,McGrew:1999nd}, 
\begin{equation}
\label{gamma-interval}
350\,\mbox{MeV}\leq q_0 \leq 600\,\mbox{MeV}\;.
\end{equation}
Since for the $2\to 2$ processes all momenta of the final particles
are fixed by the momenta of the initial particles, the above
constraint on photon frequency merely 
defines the region in the $(M_\Psi,M_\Phi)$ space where the
experimental limit \cite{Blewitt:1985zg,McGrew:1999nd} is 
applicable. 

The cross section for the process $\Psi + n\to \Phi
+\gamma$ reads  
\[
\sigma_{\Psi\,n\to\Phi\,\gamma}
= \frac{1}{64\pi M^2_nM^2_\Psi v^2} \,\overline{\left| {\cal
M}\right|^2}(t_0-t_1), 
\] 
where
\[
t_0 - t_1 = v\frac{2M_\Psi M_n}{(M_\Psi + M_n)^2}
\left((M_n+M_\Psi)^2 - M_\Phi^2\right).
\]
Finally, we obtain
\[
\sigma_{\Psi\,n\to\Phi\,\gamma}
= \frac{1}{32\,\pi\,v} \,\frac{e^2\, y^2\, \beta^2}{
4M_n^2}\frac{F^2_2(0)}{M_X^2\,\Lambda^4} \,\l 1- \frac{M_\Phi^2}{\l M_n+M_\Psi \r^2} \r\,.
\]
Similarly, the cross section of $\Phi + n\to\Psi + \gamma$ looks as
\[
\sigma_{\Phi\,n\to\Psi\,\gamma}
= \frac{1}{32\,\pi\,v} \,\frac{e^2\, y^2\, \beta^2}{
4M_n^2}\frac{F^2_2(0)}{M_X^2\,\Lambda^4} \,\l 1- \frac{M_\Psi^4}{\l
M_n+M_\Phi \r^4} \r \l 1+\frac{M_n}{M_\Phi}\r\,.
\]

\begin{figure}[!htb]
\centerline{\includegraphics[width=0.8\columnwidth,angle=-90]{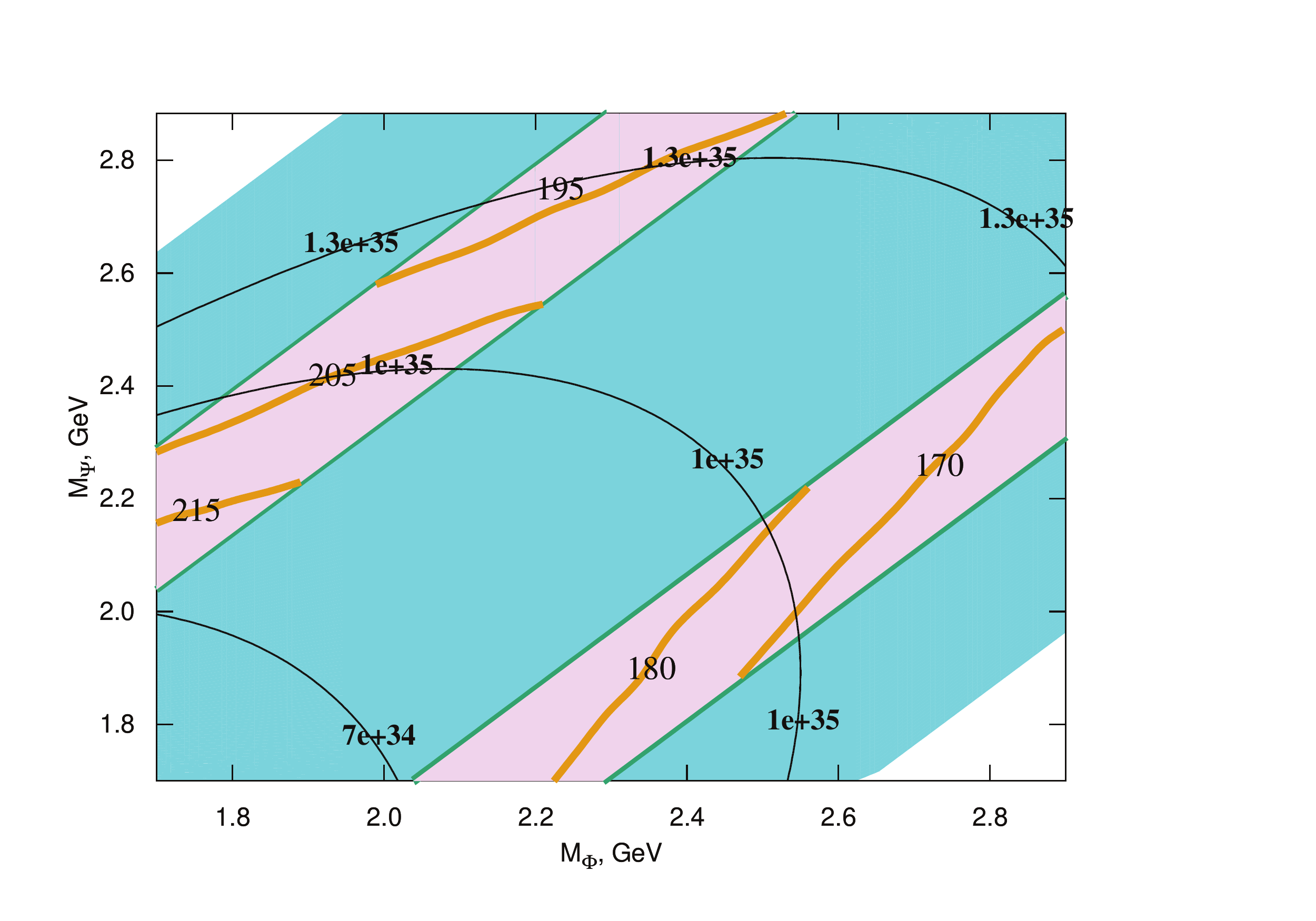} }
\caption{\label{fig:1} 
Contours (thin lines) of constant lifetime (in years) of a neutron 
with respect to the processes $\Psi n\to \Phi  \gamma$ and $\Phi
n\to \Psi  \gamma$, assuming equal number densities of the two dark
matter components and parameters $\Lambda=M_X=1$~TeV and $y=1$.
Present experimental bounds are applicable in the violet (light grey) 
regions on the plot. Thick lines in these regions show the limits on
the quantity $\left(\Lambda^2 M_X/y\right)^{1/3}$ in GeV. 
}
\end{figure}

The present lower limit on the lifetime of the neutron-decay mode in 
question is~\cite{McGrew:1999nd,Agashe:2014kda} 
\begin{equation}
\label{present-ngamma}
\tau_{n\to \gamma \nu}> 2.8\times 10^{31}\,\text{yr}\;,
\end{equation}
which is applicable in our case while the dark matter masses obey the
constraint \eqref{gamma-interval}. 
Applying eq.\,\eqref{lifetime}, in Fig.\,\ref{fig:1} 
we show contours of the constant neutron lifetime of a neutron (thin
lines) with respect to induced neutron-decay processes $\Phi(\Psi) +
n\to \Psi(\Phi) + \gamma $. They have been calculated  for the
realistic 
set of parameters $\Lambda=M_X=1$~TeV and $y=1$ without any cuts on
the phase space. As we explained above, the present experimental limit
on this process can be applied only within 
regions shown in violet (light grey) color. In this case,
one can obtain the current limit on the characteristic scale of the
process $\left(\Lambda^2 M_X/y\right)^{1/3}$; the corresponding bounds
are shown in these regions by thick lines.  Outside
shaded blue (dark grey) and violet (light grey) regions on this and
the subsequent similar plots, the stability
requirement~\eqref{kinematic-constraint}  is not satisfied. 

Note in passing, that applying LHC bounds obtained
in~\cite{Demidov:2014mda} is not quite straightforward because the
couplings $\zeta_a$ which enter~\eqref{contact} are not limited
directly from these searches. Thus, smaller values of $\Lambda$ and
$M_X$ may be allowed. However, in this paper we will use 
$\Lambda=M_X=1$~TeV and $y=1$ as a reference set of parameters for
numerical estimates.

\section{Scattering processes $\Psi(\Phi) + n\to \Phi(\Psi) +
e^+\,e^-$} 
\label{Sec:ee}

This is a $2\to3$ process induced by couplings\,\eqref{Yukawa},
and~\eqref{lagr2} through the exchange of a virtual neutron and with 
emission of virtual photon producing an electron-positron pair.  
With $p_+$ and $p_-$ being the 4-momenta of
an outcoming positron and electron, the matrix element is 
\[
\frac{ie^2y\beta }{2M_n\,\Lambda^2}\frac{F_2(q^2)}{M_X\,q^2} 
\, \overline{\Psi^C}(p_\Psi)\frac{1+\gamma_5}{2}\frac{\hat p
-M_n}{p^2-M_n^2} \sigma^{\mu\nu}q_\nu n (p_n) \Phi 
\bar\psi(p_+)\gamma^\mu \psi(p_-).
\] 
Here $q=p_++p_-$, $p = p_n-q$ and $\Psi(p_\Psi)$, $n(p_n)$,
$\psi(p_+)$, $\psi(p_-)$ are wave functions of $\Psi$, $n$, $e^+$, and
$e^-$, respectively.

For the squared matrix element of the $2\to 3$ process averaged over
spins of two incoming fermions we obtain 
\begin{equation}
\label{2to3}
\begin{split}
&\overline{\left| {\cal
M}\right|^2}=\frac{e^4\, y^2\, \beta^2}{
4M_n^2}\frac{F^2_2(q^2)}{M_X^2\,\Lambda^4} \frac{4}{q^2} 
\frac{1}{\l q^2-2qp_n\r^2} \,
\left( q^2 \cdot
p_np_+ \cdot qp_\Psi +q^2
\cdot p_+p_\Psi \cdot qp_n\right. \\
& -4 qp_\Psi\cdot
p_np_+ \cdot p_np_++4 p_np_+\cdot p_np_+\cdot p_np_\Psi+2 p_np_+\cdot
qp_n \cdot qp_\Psi 
\\
& +q^2 M_n^2 qp_\Psi -q^2 M_n^2 p_np_\Psi 
- 4 p_n p_+ \cdot qp_n \cdot p_n p_\Psi -2p_+p_\Psi \cdot qp_n\cdot qp_n\\ 
&+4 qp_n\cdot p_+ p_\Psi\cdot p_np_+ -2q^2 \cdot  p_+ p_\Psi \cdot p_np_+  +2 p_n p_\Psi \cdot qp_n \cdot
qp_n -q^2\cdot qp_n\cdot p_np_\Psi 
\left.\r.
\end{split}
\end{equation}
In what follows, it is convenient to describe the final state in terms
of energies of the outgoing visible particles by choosing, say,
positron energy $E_+$ and the sum of positron and electron energies
$E$. Then for the scattering $\Psi\, n\to \Phi\, e^+ e^-$, to the
leading order in dark matter particle velocity $v\ll 1$, one should
make the following  substitution in eq.\,\eqref{2to3} (in both the
center-of-mass and the laboratory frames)   
\begin{align*}
p_n p_\Psi&=M_nM_\Psi\,,\;\;\;p_n p_+=M_n E_+\,,\;\;\; 
qp_n=EM_n \,,\\ 
p_+p_\Psi &= E_+M_\Psi \,,\;\;\;qp_\Psi  =EM_\Psi \,,\;\;\; qp_+=EE_+\,,\\
q^2&=2EM - M^2 + M_\Phi^2\,,\;\;\;
q^2-2qp_n=2E M_\Psi - M^2 + M_\Phi^2\,,\\
\end{align*}
where we introduced the notation $M=M_n+M_\Psi$. Finally, we arrive at 
\begin{equation*}
\begin{split}
\overline{\left| {\cal
M}\right|^2}=&\frac{e^4\, y^2\, \beta^2}{
M_n^2\,q^2}\frac{F^2_2(q^2)}{M_X^2\,\Lambda^4} 
\frac{M_n\,M_\Psi}{\l q^2-2qp_n\r^2} 
\\&\times \left[
q^2 \l 2E_+(E-E_+)-M_n^2\r +2 M_n^2 \l E_+^2+(E-E_+)^2\r
\right].
\end{split}
\end{equation*}

The expression for the differential cross section looks as
follows\,\cite{Agashe:2014kda} 
\[
d\sigma = \frac{1}{4I(2\pi)^316s}\,\overline{\left| {\cal
M}\right|^2}\,dm_{12}^2\,dm_{23}^2\,,
\]
where $I=\sqrt{(p_np_{\Psi})^2-M_n^2M_\Psi^2}\approx M_nM_\Psi v$ is a
flux factor, and in the nonrelativistic limit, one has
$\sqrt{s}=M$. The 
invariant masses of outgoing pairs (let 
the subscripts ``1'' and ``2'' refer to the visible particles and
``3'' to the dark matter) in the nonrelativistic limit get reduced to  
\begin{equation}
\label{con1}
\begin{split}
m_{23}^2=M^2+m_1^2-2ME_1\,,\;\;&m_{12}^2=2ME-M^2+m_3^2 \\
dm_{12}^2 \, dm_{23}^2 = - & 2M dE \, 2M dE_1.
\end{split}
\end{equation}
The energy $E$ is confined within the interval 
\begin{equation}
\label{con2}
\frac{(m_1+m_2)^2+M^2-m_3^2}{2M}<E<M-m_3
\end{equation}
and $E_1$ is within the interval
\begin{equation}
\label{con3}
\frac{m_1^2+M^2-(m_{23}^2)^{max}}{2M}<E_1<\frac{M^2+m_1^2-(m_{23}^2)_{min}}{2M}, 
\end{equation}
where 
\[
(m_{23}^2)^{max}_{min}=2E_2^*E_3^*-m_2^2-m_3^2\pm
2\sqrt{E_2^{*2}-m_2^2}\sqrt{E_3^{*2}-m_3^2} 
\]
and 
\begin{align}
E_3^{*2}-m_3^2&=\frac{M^2((M-E)^2-m_3^2)}{2ME-M^2+m_3^2}\,, \\
E_2^{*2}-m_2^2&=\frac{(2ME-M^2+m_3^2+m_2^2-m_1^2)^2}{4(2ME-M^2+m_3^2)}-m_2^2\,,\\
\label{6.6}
2E_2^*E_3^*&=\frac{(-M^2+2ME+m_3^2+m_2^2-m_1^2)(M^2-ME-m_3^2)}{2ME-M^2+m_3^2}\,. 
\end{align}

For the process under discussion, let subscript ``1'' refer to the
positron, and replacing in the above formulas $E_1$ with $E_+$, we
obtain the differential cross section 
\[
d\sigma = \frac{1}{128\pi^3vM_nM_\Psi}\,\overline{\left| {\cal
M}\right|^2}\,dE\,dE_+
\]
which must be integrated over the region defined by
eqs.\,\eqref{con1}--\eqref{6.6}. 

For the process $\Phi(p_\Phi)\,
n(p_n) \to \Psi(p_\Psi)\, e^+(p_+) e^-(p_-)$, one has the same
expression \eqref{2to3} multiplied by a factor of 2 due to one less
number of initial fermions and makes the replacement  
\begin{align*}
p_np_+ & = M_nE_+\;,\;\;\; qp_n = M_nE\;, \;\;\;
p_np_\Psi  =  M_n(M-E)\,, \\ q^2 & = - M^2 + M_\Psi^2 + 2ME\;,\;\;\;
p_+p_\Psi = \frac{1}{2}M^2 - \frac{1}{2}M_\Psi^2 - M(E-E_+)\;,\\
qp_\Psi&=M^2-M_\Psi^2-ME\,,\;\;\;
q^2-2qp_n = -M^2+M_\Psi^2+2E(M-M_n)\;,
\end{align*}
where $M = M_\Phi + M_n$.

The current best limit\,\cite{Agashe:2014kda,McGrew:1999nd} for
neutron decay in the mode $n\to\nu e^+e^-$ is: 
\[
\tau_{n\to\nu e^+e^-}> 2.57\times10^{32}\,\text{yr}\;. 
\]
It has been obtained from the analysis of experimental data with imposing
the following cut on the total energy of leptons\,\cite{Blewitt:1985zg,McGrew:1999nd}  
\begin{equation}
\label{cut1}
500\,\text{MeV}\leq E \leq 850\,\text{MeV}
\end{equation}
and assuming that the asymmetry is small, 
\begin{equation}
\label{cut2}
A<0.5\,.
\end{equation} 
The latter quantity characterizes the directional asymmetry of energy
release in the Cherenkov detector. The asymmetry is maximal, $A=1$,
for collinear particles and equals zero for such a decay,  where
the particles go in opposite directions. Let us stress, that this
quantity counts not all the particles, but only those which release
the energy inside the Cherenkov detector, and accounts for them with 
weights proportional to the energy release into the Cherenkov
radiation.   

In our case of the electron-positron pair the weights are identical.  
For the decay $n\to \nu\, e^+ \,e^-$,
all 3-momenta of the outgoing particles are in a decay plane. All 
three particles are relativistic, so the Cherenkov angles for the
electron and positron are identical and the energy conservation gives
for the sum of the particle energies
\begin{equation}
\label{energy-conservation}
E_\nu+E_++E_-=M \,,
\end{equation}
where $M$ is the neutron mass. 
Then the {\it asymmetry} defined in\,\cite{Blewitt:1985zg,McGrew:1999nd}  is just 
\begin{equation}
\label{asymmetry-via-directions}
A\equiv \half \l 1+ {\bf n_+} {\bf n_-} \r,
\end{equation}
where ${\bf n_\pm}$ are unit 3-vectors along the direction of the
outgoing positron and electron, respectively. Introducing the
reference axis along the 3-momentum of the neutrino, one defines
corresponding transverse and longitudinal parts of the electron and 
positron momenta. Obviously, the transverse parts of electron and
positron momenta are equal in magnitude but of opposite directions 
\begin{equation}
\label{perp}
p_+^{\perp}=-p_-^{\perp},
\end{equation}
while the longitudinal parts (momentum projection on the chosen axis)
sum to zero, 
\begin{equation}
\label{parallel}
p_\nu^{\parallel}+p_+^{\parallel}+p_-^{\parallel}=0.
\end{equation}
For the relativistic electron and positron, one has 
\begin{equation}
\label{magnitudes}
E_\pm^{2}=p_\pm^{\parallel\,2}+p_\pm^{\perp\,2}
\end{equation}
and for relativistic neutrino with the chosen axis
$p_\nu^{\parallel}>0$ and 
\begin{equation}
\label{neutrino-energy}
p_\nu^{\parallel}=E_\nu.
\end{equation}
Then, the asymmetry \eqref{asymmetry-via-directions} reads
\begin{equation}
\label{asymmetry-via-momenta}
A=\half \l 1 + \frac{p_+^{\parallel}}{E_+}\frac{p_-^{\parallel}}{E_-} 
+ \frac{p_+^{\perp}}{E_+}\frac{p_-^{\perp}}{E_-} \r.
\end{equation}

The differential decay rate is given by 
 \begin{equation}
\label{decay-rate}
d\Gamma = \frac{1}{(2\,\pi)^3}\, \frac{1}{32\,M^3}\, 
\overline{\left| {\cal M}\right|^2}\, dm_{12}^2\, dm_{23}^2\,. 
\end{equation}
Introducing the sum of the electron and positron energies 
\[
E\equiv E_++E_-
\]
one obtains for the phase space measure \eqref{con1} (where $E_1$
stands for $E_+$) that ranges \eqref{con2} and \eqref{con3} are
reduced to  
\begin{equation}
\label{phase-space}
\frac{M}{2}<E<M\,\;\;\;\;\;E-\frac{M}{2}<E_+<\frac{M}{2}\;.
\end{equation}
Two independent variables, e.g. $E$ and $E_+$, fix all the others,
which can be found by solving 
Eqs.\,\eqref{energy-conservation}, \eqref{perp}, \eqref{parallel},
\eqref{magnitudes} under condition \eqref{neutrino-energy}. The
results read
\begin{align}
E_-&=E-E_+\,,\\
E_\nu&=M-E\,,\\
p_+^{\parallel}&= \frac{E(M-E_+)-M^2/2}{M-E}\,,\\
p_-^{\parallel}&= \frac{E(M-E)+EE_+-M^2/2}{E-M}\,,\\
p_-^{\perp\,2}&=p_+^{\perp\,2}=\frac{M(E-M/2)(2E_+-M)(E-E_+-M/2)}{(E-M)^2}.
\end{align}
 
Putting the solutions above into \eqref{asymmetry-via-momenta}, one
obtains for the asymmetry 
\begin{equation}
\label{asymmetry-via-energy}
A=\half \l 1 - \frac{E_+(E_+-E)+M(E-M/2)}{E_+(E-E_+)}\r.
\end{equation}
The cut adopted in \cite{Blewitt:1985zg,McGrew:1999nd} $A<0.5$ implies
a positive value of the second term in parentheses in
Eq.\,\eqref{asymmetry-via-energy}. It slightly increases the lower
limit for $E$ and, thus, reduces a little the triangle integration
region in \eqref{phase-space}.
 
To adopt the same cuts on asymmetry $A$ in the case of the $2\to 3$
process $\Psi n\to \Phi e^+ e^-$ one can treat it in the
nonrelativistic regime as a decay of the particle of effective mass  
\[
M \approx M_n+M_\Psi\,. 
\]

Then the following formulas from the previous considerations must be
modified as follows:
\begin{itemize}
\item Instead of the massless neutrino, the outcoming dark matter
particle $\Phi$ is massive, so its 3-momentum (we use the same
notations) instead of \eqref{neutrino-energy}
obeys 
\begin{equation}
\label{neutrino-energy-bis}
p_\nu^{\parallel\,2}+M^2_\Phi=E_\nu^2,
\end{equation}
\item The region of integration in Eq.\,\eqref{phase-space}
\begin{equation}
\label{phase-space-bis}
\frac{M^2-M^2_\Phi}{2M}<  E <M-M_\Phi  
\,, \;\; 
\end{equation}
\[
\frac{1}{2}(E-\sqrt{(M-E)^2-M_\Phi^2}) < E_+ < \frac{1}{2}(E+\sqrt{(M-E)^2-M_\Phi^2}) ,
\]
\item Longitudinal momenta are 
\begin{align}
p_+^{\parallel}&= \frac{E(M-E_+)-M^2/2+M_\Phi^2/2}{\sqrt{(M-E)^2-M_\Phi^2}}\,,\\
p_-^{\parallel}&= \frac{E(E_++M-E)-M^2/2+M_\Phi^2/2}{\sqrt{(M-E)^2-M_\Phi^2}}\,,
\end{align}
and the transverse momenta read
\begin{equation}
p_-^{\perp\,2}=
p_+^{\perp\,2}=
\frac{[M(E_+-E+M/2)-M_\Phi^2/2][(E-M/2)(M-2E_+)+M_\Phi^2/2]}{(M-E)^2-M_\Phi^2}  ,
\end{equation}
\item 
The asymmetry \eqref{asymmetry-via-energy} must be replaced with 
\begin{equation}
\label{asymmetry-via-energy-bis}
A=\half \l 1 - \frac{E_+(E_+-E)+M(E-M/2)+M_\Phi^2/2}{E_+(E-E_+)}\r.
\end{equation}
\end{itemize}

Similar formulas with evident replacements $M_\Phi\to M_\Psi$ and
$M\to M_n+ M_\Phi$ are applicable for the description of the twin
process $\Phi n\to \Psi e^+e^-$.  

\begin{figure}[!htb]
\centerline{\includegraphics[width=0.8\columnwidth,angle=-90]{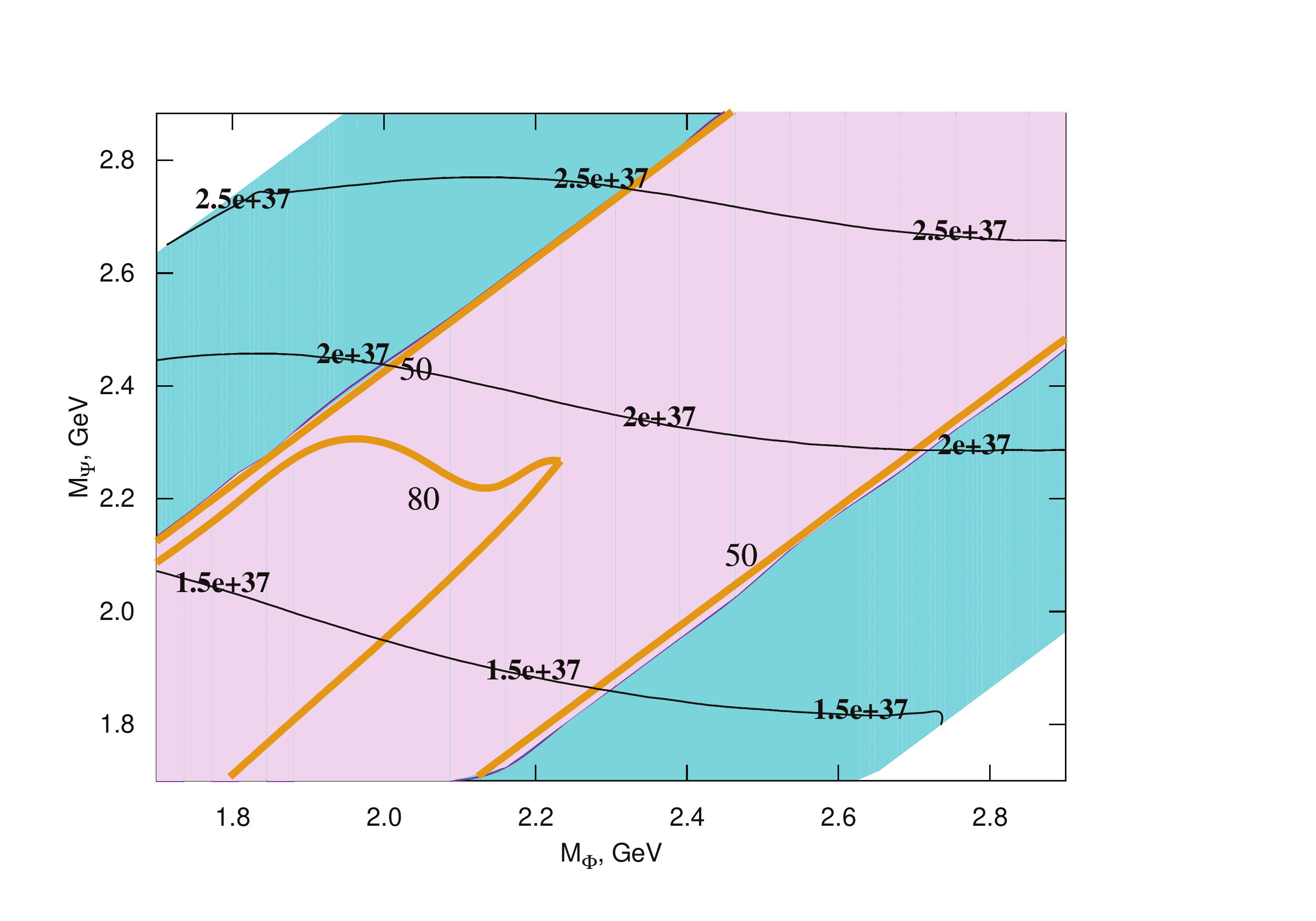}}
\caption {\label{fig:2}
Contours (thin lines) of constant lifetime (in years) of a neutron
with respect to the process  $\Psi(\Phi) n\to \Phi(\Psi)  e^+e^-$; 
we set $\Lambda=M_X=1$~TeV and $y=1$.  
Present experimental bounds are applicable in the violet (light grey)
region on the plot. Thick lines in this region show the limits on
the quantity $\left(\Lambda^2 M_X/y\right)^{1/3}$.
} 
\end{figure}

For the original process $n\to \nu e^+e^-$, assuming the
momenta-independent matrix element, the cuts \eqref{cut1} and
\eqref{cut2} select a 0.3278/0.3904 part 
of the phase space. 
In Fig.\,\ref{fig:2},
we show contours of the constant lifetime of a neutron (thin lines)
with respect to induced neutron decay processes $\Phi(\Psi) +
n\to \Psi(\Phi) + e^+e^- $. They have been calculated without any cuts
for $\Lambda=M_X=1$~TeV and $y=1$. The current limits on these
processes can be applied only within the region shown in violet (light 
grey) color. They are distinguished by the corresponding kinematics of
the process and applied cuts \eqref{cut1}, \eqref{cut2}. In this case
one can obtain the current limit on the characteristic scale of the
process $\Lambda$; the corresponding bounds are shown in these regions
by thick lines. 

\begin{figure}[!htb]
\centerline{\includegraphics[width=0.7\columnwidth,angle=-90]{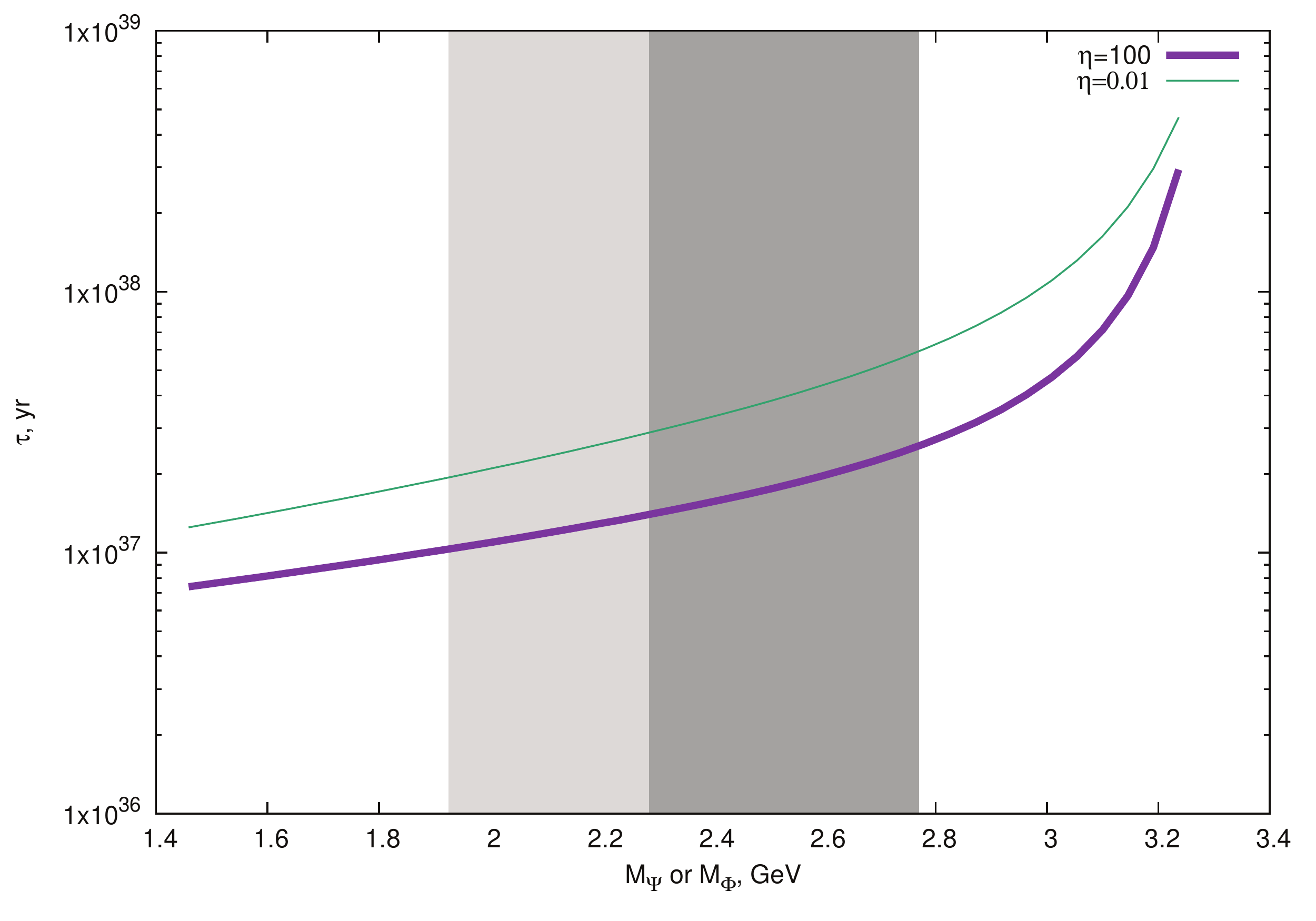} }
\caption{\label{fig:3} Lifetime of a neutron with respect to IND $n\to
e^+e^-$ for $\eta=100$ and  $\eta=0.01$; we set $\Lambda=M_X=1$~TeV
and $y=1$. The current limits are applied in shaded regions.} 
\end{figure}

Now let us consider the asymmetric case when number densities of
$\Psi$ and $\Phi$ are different, $\eta\neq 1$, see eq.\,\eqref{ratio}.  
As an example, below we consider opposite cases of asymmetry:
$\eta=0.01$ and $\eta=100$, which correspond to $\Psi$ or $\Phi$
dominance, respectively. Note that in this case the allowed mass
intervals are different from that of the symmetric case. Namely, mass
of the dominant component is fixed in the very narrow region around
$5M_p/2$, while the subdominant component can have mass which is
determined by the condition \eqref{kinematic-constraint}. 
In Fig.~\ref{fig:3}, 
we show expected lifetimes of a neutron with respect to the processes
$\Phi(\Psi) + n\to \Psi(\Phi) + e^+e^- $ for the cases of $\Phi$ and
$\Psi$ dominance calculated for the same set of parameters as we
described previously. The current limits on this process are
applicable in the shaded regions on these figures and they (almost
uniformly over these regions) result in $\Lambda>35$~GeV ($\eta=100$)
and $\Lambda> 40$~GeV ($\eta=0.01$) for region 1 and $\Lambda>86$~GeV
($\eta=100$) and $\Lambda>75$~GeV ($\eta=0.01$) for region 2.

\begin{figure}[!htb]
\centerline{\includegraphics[width=0.7\columnwidth,angle=-90]{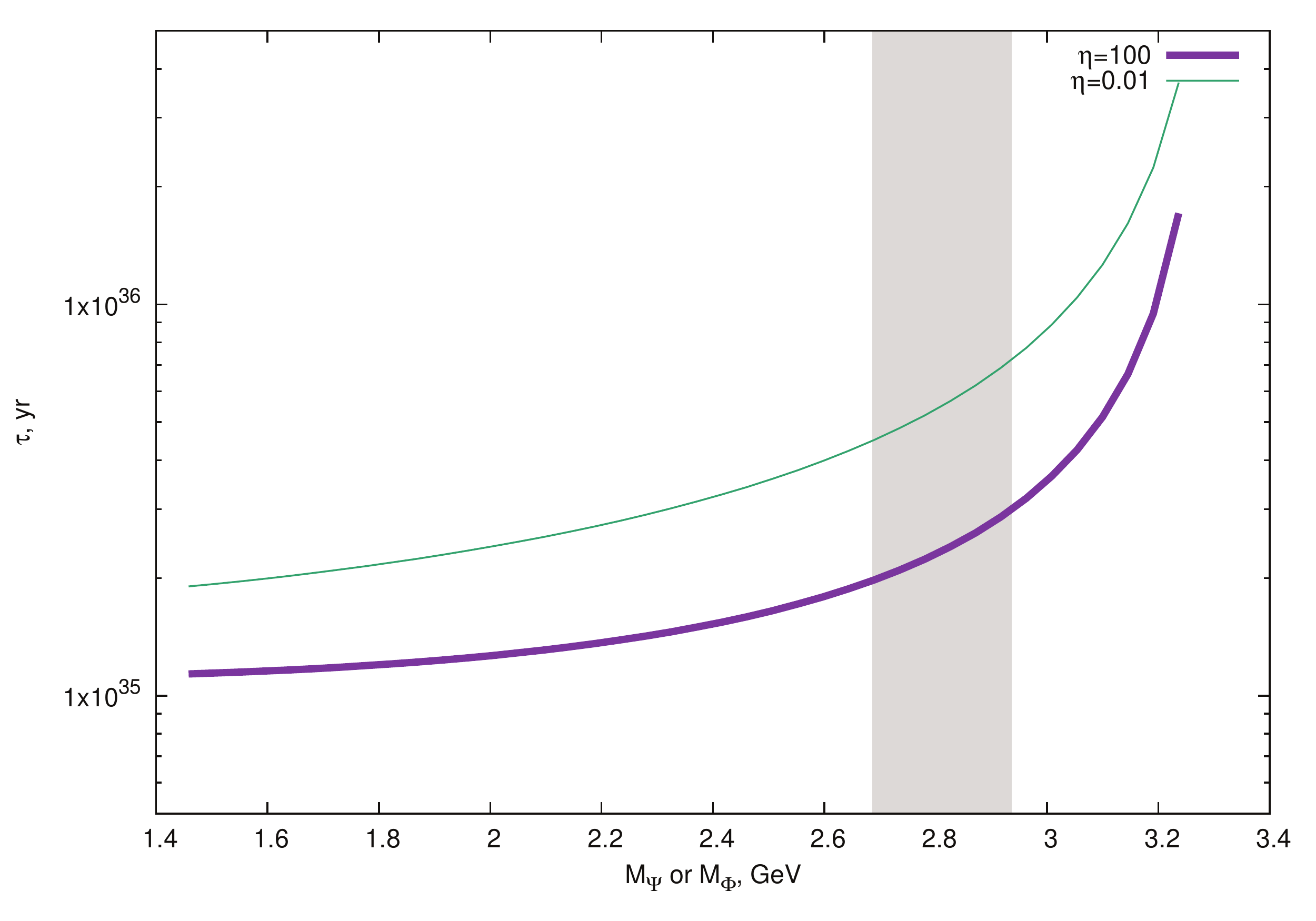} }
\caption{\label{fig:4} Lifetime of a neutron with respect to the
processes $\Phi + n\to \Psi + \gamma $ (for $\eta=100$) and $\Psi +
n\to \Phi + \gamma $ (for $\eta=0.01$); we set $\Lambda=M_X=1$~TeV and
$y=1$. } 
\end{figure}

Similar plots for the processes $\Phi(\Psi) + n\to \Psi(\Phi)
+ \gamma $ are shown in Fig.~\ref{fig:4}. Here one
can obtain the following limits on $\Lambda$: for $\eta=100$, we have
$\Lambda > 215-228$~GeV, and for  $\eta=0.01$, we obtain  $\Lambda >
185-200$~GeV depending on the mass of the subdominant component. 

\section{Processes $\Psi(\Phi) + N \to \Phi(\Psi) + 2$ mesons}
\label{Sec:2mesons}

Within the chiral perturbation theory, the IND processes with two
mesons in the final state arise in the $1/f^2$ order due to the
following terms in the low-energy effective lagrangian  
\begin{equation}
\label{lagr0}
{\cal L}_{1\pi} = i\,\frac{c_1\beta}{f}\,\Phi\overline{\Psi^C}
\l-\sqrt{\frac{3}{2}}n\eta + \frac{1}{\sqrt{2}}n\pi^0 -
p\pi^-\r + \text{h.c.}\,,
\end{equation}
\begin{equation}
\label{lagr1}
\begin{split}
{\cal L}_{2\pi} = -\frac{\beta\,c_1}{2f^2} 
\l\sqrt{6}\pi^-\eta + K^0{K}^-\r \Phi\,\overline{\Psi^C}p_R\\ 
-\frac{\beta\,c_1}{2f^2}  \l\pi^+\pi^- + \frac{3}{2}\eta^2
- \sqrt{3}\eta\pi^0 + \frac{1}{2}(\pi^{0})^2 + 2K^0\bar{K}^0
+ K^+K^-\r \Phi\,\overline{\Psi^C} n_R +\text{h.c.}\,,
\end{split}
\end{equation}
with parameter $c_1$ related to the model parameters as follows from  
matching eqs.\,\eqref{contact} and \eqref{new-variables} to 
eqs.\,\eqref{appA:1} and \eqref{appA:2}
\[
c_1 = \frac{y}{M_X\Lambda^2}\,.
\]
Details of the derivation are presented in the Appendix\,\ref{App:A} 
for completeness. Below, we work in the limit of exact isotopic
invariance, neglecting the proton-neutron and charged-neutral pion
mass differences,  
\[
M_n=M_p\equiv M_N\,,\;\;\;\; m_{\pi^+}=m_{\pi^0}\equiv
m_\pi\,,\;\;\;\; m_{K^+}=m_{K^0}\equiv m_K\;.
\]  

Two types of diagrams contribute the processes: one of them follows 
from lagrangian~\eqref{lagr1} and the other comes from one-meson
lagrangian~\eqref{lagr0}, while the second meson is radiated from the
nucleon leg; see eq.\,\eqref{App:lagr_nucl}.

For the $dud$ operator, we have the following possibilities for
induced decays, which we classify here according to the number of
tree-level Feynman diagrams contributing the corresponding 
process:
\begin{itemize}
\item one-diagram processes $p\to \bar{K}^0K^+$,  $n\to K^0\bar{K}^0$,
and $n\to K^+K^-$;
\item two-diagram processes $p\to\pi^0\pi^+$, $n\to\pi^+\pi^-$;
\item three-diagram processes $n\to\eta \pi^0$, $p\to\eta\pi^+$,
$n\to\eta\eta$, and $n\to \pi^0\pi^0$.
\end{itemize}

\begin{figure}[!htb]
\centerline{\includegraphics[width=0.3\columnwidth]{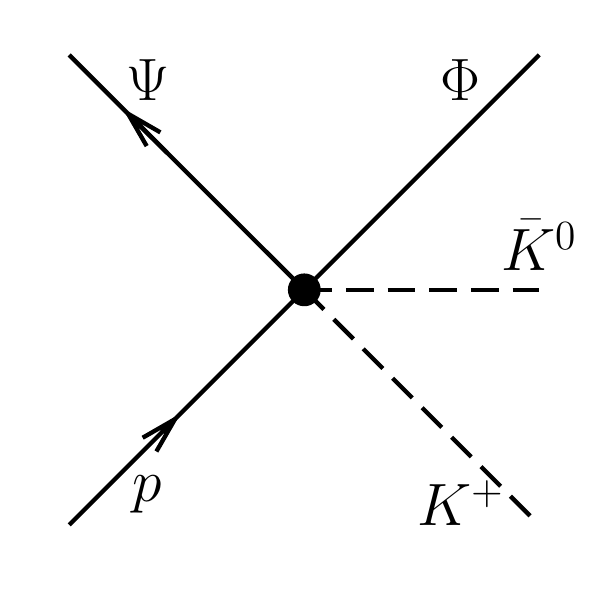} }
\caption{\label{fig_Feyn1} The Feynman diagrams for the process $\Psi+p\to \Phi+\bar{K}^0 K^+$ .} 
\end{figure}

\paragraph{One-diagram processes.}
The Feynman diagram for the process $\Psi+p\to \Phi+\bar{K}^0 K^+$ is
presented in Fig.~\ref{fig_Feyn1}
Averaged over spins of the two initial fermions, the squared matrix
element of this process reads 
\begin{equation}
\label{KK}
\overline{\left| {\cal M}\right|^2}=\frac{\beta^2 c^2_1}{8
f^4}\,p_N p_\Psi\,,
\end{equation}
and by a factor of two bigger for $\Phi+p\to \Psi+\bar{K}^0 K^+$; 
hereafter, $p_N$ refers to the 4-momentum of the nucleon participating
in the corresponding process. For the first process, in
the laboratory frame one has to the leading order in dark matter velocity
\begin{equation}
\label{1d-decay}
p_Np_\Psi=M_N\,M_\Psi
\end{equation}
and when integrating over the phase space adopts the formulas 
\eqref{con1}--\eqref{6.6}   with 
\begin{equation}
\label{1d-decay-2}
M=M_N+M_\Psi\;,\;\;m_1=m_2=m_K\,,\;\;\;m_3=M_\Phi\,. 
\end{equation}
Instead, for the second process we have
\begin{equation}
\label{1d-decay-3}
M=M_N+M_\Phi\,,\;\;p_Np_\Psi=M_N\,\l M -E\r\,,\;\;
m_1=m_2=m_K\,,\;\;m_3=M_\Psi. 
\end{equation}

Averaged over spins of the initial two fermions, squared matrix
element of the process $\Psi+n\to \Phi+K^- K^+$ reads as \eqref{KK} 
and by a factor of two bigger for $\Phi+n\to \Psi+K^- K^+$. Further,
in the laboratory frame one can use
eqs.\,\eqref{1d-decay}, \eqref{1d-decay-2} and eq.\,\eqref{1d-decay-3}
for the first and second processes, respectively. The same sets of 
formulas work for the processes 
$\Psi+n\to \Phi+\bar{K}^0 K^0$ and $\Phi+n\to \Psi+\bar{K}^0 K^0$, 
respectively.

\paragraph{Two-diagram processes.}

To describe this class of processes it is convenient to introduce the
following notations 
\begin{align}
I(p_1,p_2) &\equiv 2\,p_1p_2-p_2^2\,,\\
J(p_1,p_2,p_3) &\equiv 2\,p_1p_3\cdot p_2p_3 - p_3^2\cdot p_1p_2\,,
\\
K(p_1,p_2,p_3,p_4) &\equiv p_1p_3\cdot p_2p_4 + p_1p_4\cdot p_2p_3 
- p_1p_2\cdot p_3p_4\,.
\end{align}

\begin{figure}[!htb]
\centerline{\includegraphics[width=0.7\columnwidth]{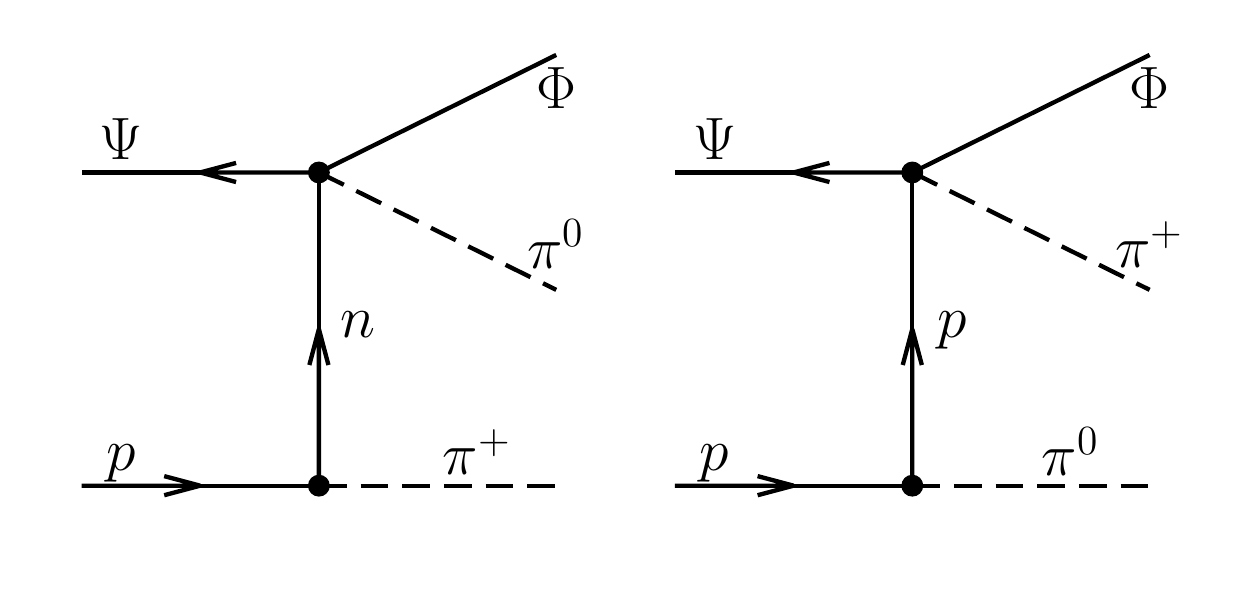} }
\caption{\label{fig_Feyn2} The Feynman diagrams for the process $\Psi+p\to \Phi+\pi^+\pi^0$.} 
\end{figure}

The Feynman diagrams for the process $\Psi+p\to \Phi+\pi^+\pi^0$ are 
presented in Fig.~\ref{fig_Feyn2}.
The squared matrix element of this process,
averaged over spins of initial particles is        
\begin{equation}
\label{2d-decay-1}
\begin{split}
\overline{\left| {\cal M}\right|^2}=&
\frac{(D+F)^2c_1^2\beta^2M_N^2}{f^4} \\&\times \l
\frac{J(p_N,p_\Psi,p_{\pi^+})}{I^2(p_N,p_{\pi^+})}
+ \frac{J(p_N,p_\Psi,p_{\pi^0})}{I^2(p_N,p_{\pi^0})}
-\frac{2K(p_N,p_{\Psi},p_{\pi^+},p_{\pi^0})}
{I(p_N,p_{\pi^+})\,I(p_N,p_{\pi^0})}\r,
\end{split}
\end{equation}
where $D=0.8$ and $F=0.47$ (see the Appendix~A). In the laboratory
frame, one has  
\begin{align*}
p_Np_\Psi & =M_N M_\Psi\,,\;\;\;
p_Np_{\pi^0}=M_N(E-E_1),\;\;\;p_N p_{\pi^+} = M_N E_1\,,\\ 
p_{\pi^0}p_\Psi &=\l E-E_1\r M_\Psi\,,\;\;\;
p_{\pi^+}p_\Psi =E_1 M_\Psi\,,\;\;\;
p_{\pi^+}p_{\pi^0} =E M -m_\pi^2+\frac{1}{2}\l
M_\Phi^2-M^2\r\,,
\end{align*}
and adopts eqs.\eqref{con1}--\eqref{6.6} with 
\[
M=M_\Psi+M_N, \;\; m_1=m_2=m_\pi\,,\;\;\;m_3=M_\Phi\,.
\]

\begin{figure}[!htb]
\centerline{\includegraphics[width=0.8\columnwidth,angle=-90]{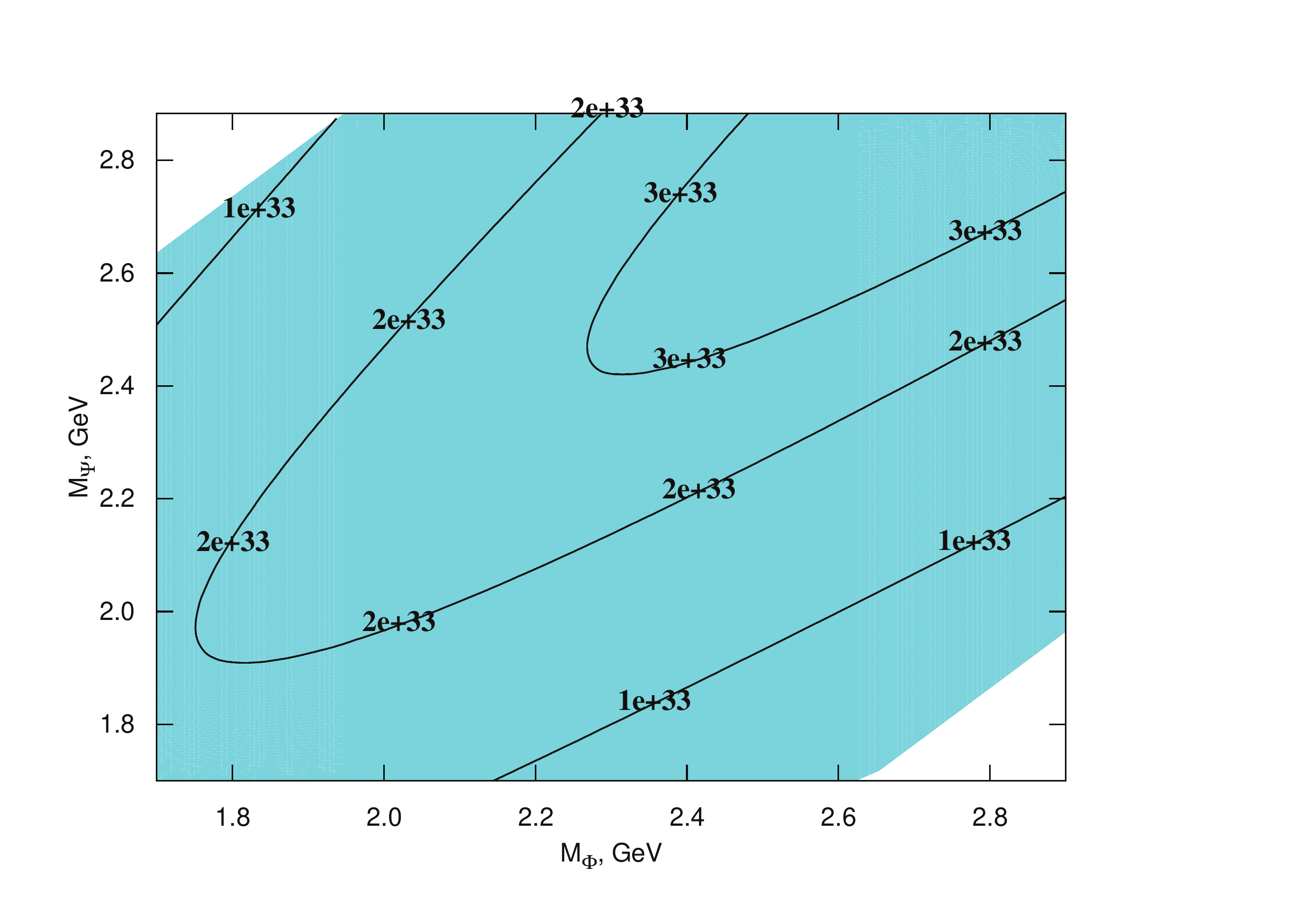} }
\caption{\label{fig:5} Contours of constant lifetime (in years) 
of the nucleon in the symmetric case with respect 
to IND process with $\pi^+\pi^0$ in the final state; we set
$\Lambda=M_X=1$~TeV and $y=1$.} 
\end{figure}

For $\Phi+p\to \Psi+\pi^+\pi^0$, one obtains for the squared averaged
matrix element \eqref{2d-decay-1} but a factor of two bigger. In the
laboratory frame, one finds 
\begin{align*}
p_Np_\Psi & =M_N \l M-E\r\,,\;\;
p_{\pi^0}p_\Psi =\frac{1}{2}\l
M^2-M_\Psi^2\r -E_1 M
\,,\\ 
p_Np_{\pi^0}&=M_N(E-E_1)\,,\;\;p_{\pi^+}p_{\pi^0} =
EM -m_\pi^2+\frac{1}{2}\l
M_\Psi^2-M^2\r
\,,\\
p_N p_{\pi^+} &= M_N E_1\,,\;\;
p_{\pi^+}p_\Psi =\frac{1}{2}\l
M^2-M_\Psi^2\r - \l E-E_1\r M\,,
\end{align*}
with 
\[
M=M_\Phi+M_N\,,\;\;\; m_1=m_2=m_\pi\,,\;\;\;m_3=M_\Psi\,.
\]
The predictions of the proton decay with $\pi^+\pi^0$ final state are
presented in Fig.\,\ref{fig:5} as contours of the constant lifetime
for the symmetric case $\eta=1$. Again, here and below we fix
$\Lambda=M_X=1$~TeV and $y=1$ and impose no cuts in the phase
space.

\begin{figure}[!htb]
\centerline{\includegraphics[width=0.6\columnwidth]{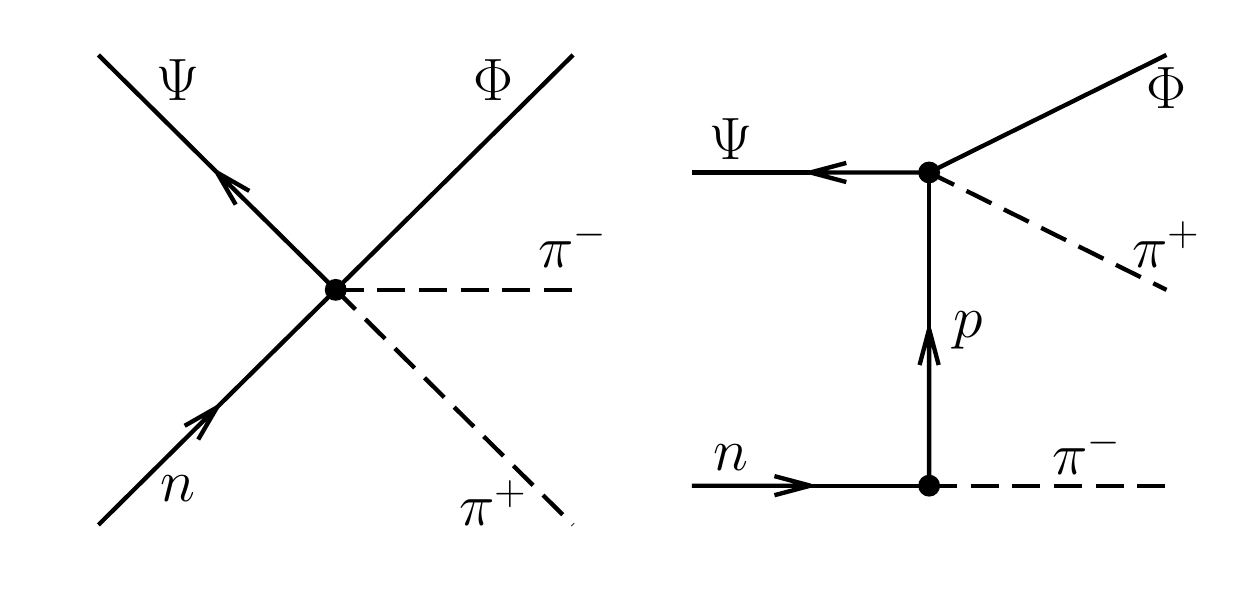} }
\caption{\label{fig_Feyn3} The Feynman diagrams for the process $\Psi+n\to \Phi+\pi^+\pi^-$ .} 
\end{figure}

Another two-diagram IND process is $\Psi+n\to \Phi+\pi^+\pi^-$.
Corresponding Feynman diagrams  are  presented in
Fig.~\ref{fig_Feyn3}. 
The squared matrix element of 
$\Psi+n\to \Phi+\pi^+\pi^-$ averaged over spins of the initial
particles takes the form 
\[
\frac{1}{2}(A-B)^2\,p_Np_\Psi + 
2B(A-B)\,\frac{M_N^2\,p_\Psi p_{\pi^-}}{I(p_N,p_{\pi^-})}
+2B^2M_N^2\frac{J(p_N,p_\Psi,p_{\pi^-})}{I^2(p_N,p_{\pi^-})}.
\]
where 
\[
A = \frac{\beta c_1}{2f^2}\,,\;\;\; B = \frac{(F+D)\beta c_1}{f^2}.
\]
In the laboratory frame, one has the same expression
as \eqref{3d-decay-1a}, \eqref{3d-decay-1b} for pions.   
For $\Phi+n\to \Psi+\pi^+\pi^-$, one has a factor of two bigger squared
averaged matrix element and the same expressions in the laboratory
frame as \eqref{3d-decay-2a} and \eqref{3d-decay-2b} for pions.

\begin{figure}[!htb]
\centerline{\includegraphics[width=0.9\columnwidth]{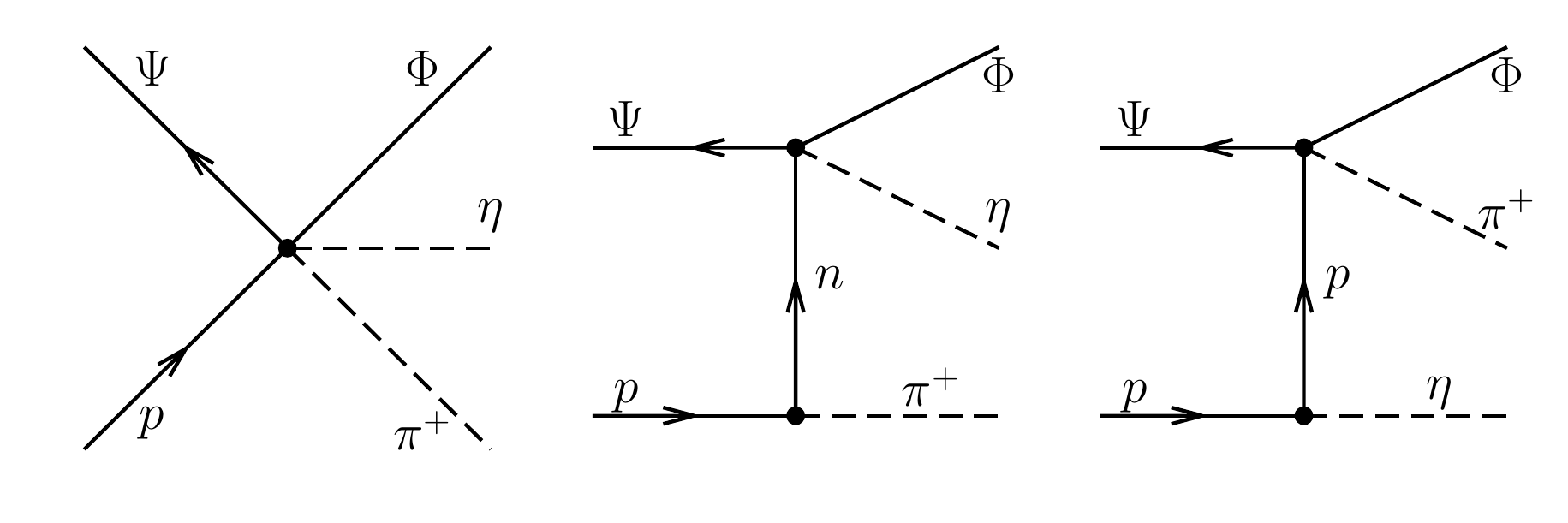} }
\caption{\label{fig_Feyn4} The Feynman diagrams for the process $\Psi+p\to \Phi+\pi^+\eta$ .} 
\end{figure}

\paragraph{Three-diagram processes.}
The Feynman diagrams for the process $\Psi+p\to \Phi+\pi^+\eta$  are 
presented in Fig.~\ref{fig_Feyn4}. 
The squared matrix elements of the processes 
$\Psi+p\to \Phi+\pi^+\eta$ and $\Psi+n\to \Phi+\pi^0\eta$,
averaged over spins of the initial particles have the form
\begin{align*}
&\overline{\left| {\cal M}\right|^2}=
2M_N^2\l B^2\frac{J(p_N,p_\Psi,p_{\pi^+})}{I^2(p_N,p_{\pi^+})}
+ C^2\frac{J(p_N,p_\Psi,p_{\eta})}{I^2(p_N,p_{\eta})}
+2BC\frac{K(p_N,p_\Psi,p_{\pi^+},p_{\eta})}
{I(p_N,p_{\pi^+})\,I(p_N,p_\eta)}\r \\ 
&+\frac{1}{2}(A-B-C)^2\,p_Np_\Psi
+2B(A-B-C)\frac{M_N^2\,p_{\pi^+}p_\Psi}{I(p_N,p_{\pi^+})}
+2C(A-B-C)\frac{M_N^2\,p_{\eta}p_\Psi}{I(p_N,p_{\eta})},
\end{align*}
where
\begin{equation}
A = \frac{\sqrt{6}\beta c_1}{2f^2},\;\;\;
B = \sqrt{\frac{3}{2}}\frac{(D+F)\beta c_1}{f^2},\;\;\;
C = \frac{(3F-D)\beta c_1}{\sqrt{6}f^2}
\end{equation}
for $\Psi+p\to \Phi+\pi^+\eta$ and
\begin{equation}
A = \frac{\sqrt{3}\beta c_1}{2f^2},\;\;\;
B = \frac{\sqrt{3}(D+F)\beta c_1}{2f^2},\;\;\;
C = \frac{(3F-D)\beta c_1}{2\sqrt{3}f^2}
\end{equation}
for $\Psi+n\to \Phi+\pi^0\eta$. In the laboratory frame, one has 
\begin{align*}
p_Np_\Psi & =M_N M_\Psi\,,\;\;\;
p_\Psi p_{\eta}=M_\Psi(E-E_1)\,,\;\;\;\,p_N p_{\pi^+} = M_N E_1\,,\\ 
p_Np_\eta &=\l E-E_1\r M_N\,,\;\;
2\,p_{\pi^+}p_{\eta} =M \l 2E-M\r+
M_\Phi^2-m_\pi^2-m_\eta^2\,,\;\;
p_{\pi^+}p_\Psi =E_1 M_\Psi\,,
\end{align*}
and utilizes eqs.\,\eqref{con1}--\eqref{6.6} with 
\[
M=M_\Psi+M_N\,\;\;\;m_1=m_\pi\,,\;\;\;m_2=m_\eta\,,\;\;\;m_3=M_\Phi\,.
\]

\begin{figure}[!htb]
\centerline{\includegraphics[width=0.8\columnwidth,angle=-90]{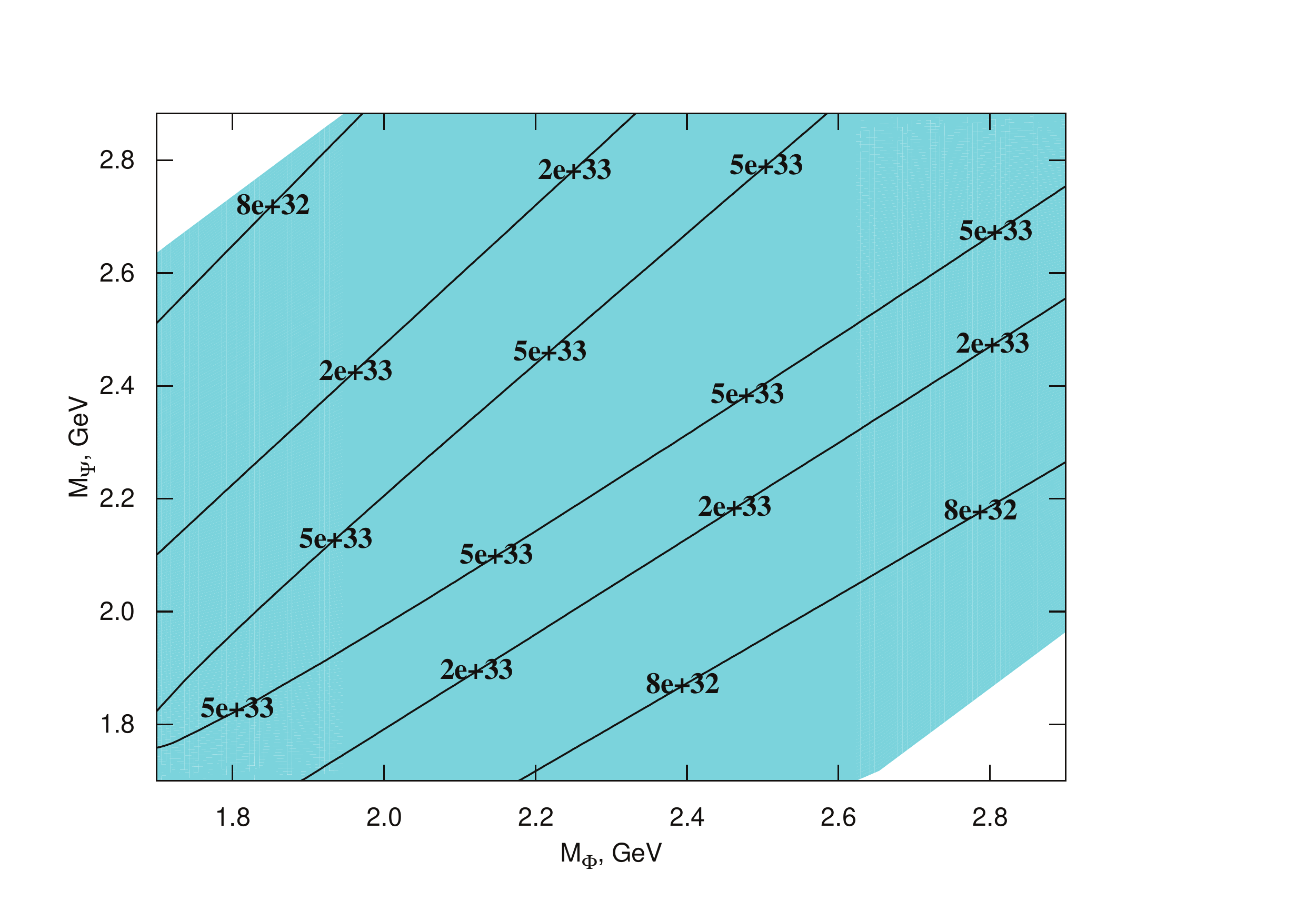} }
\caption{\label{fig:7} Contours of the constant lifetime (in years) of
a nucleon in the symmetric case with respect to IND with $\pi^+\eta$
in the final state; we set $\Lambda=M_X=1$~TeV and $y=1$.} 
\end{figure}

\begin{figure}[!htb]
\centerline{\includegraphics[width=0.8\columnwidth,angle=-90]{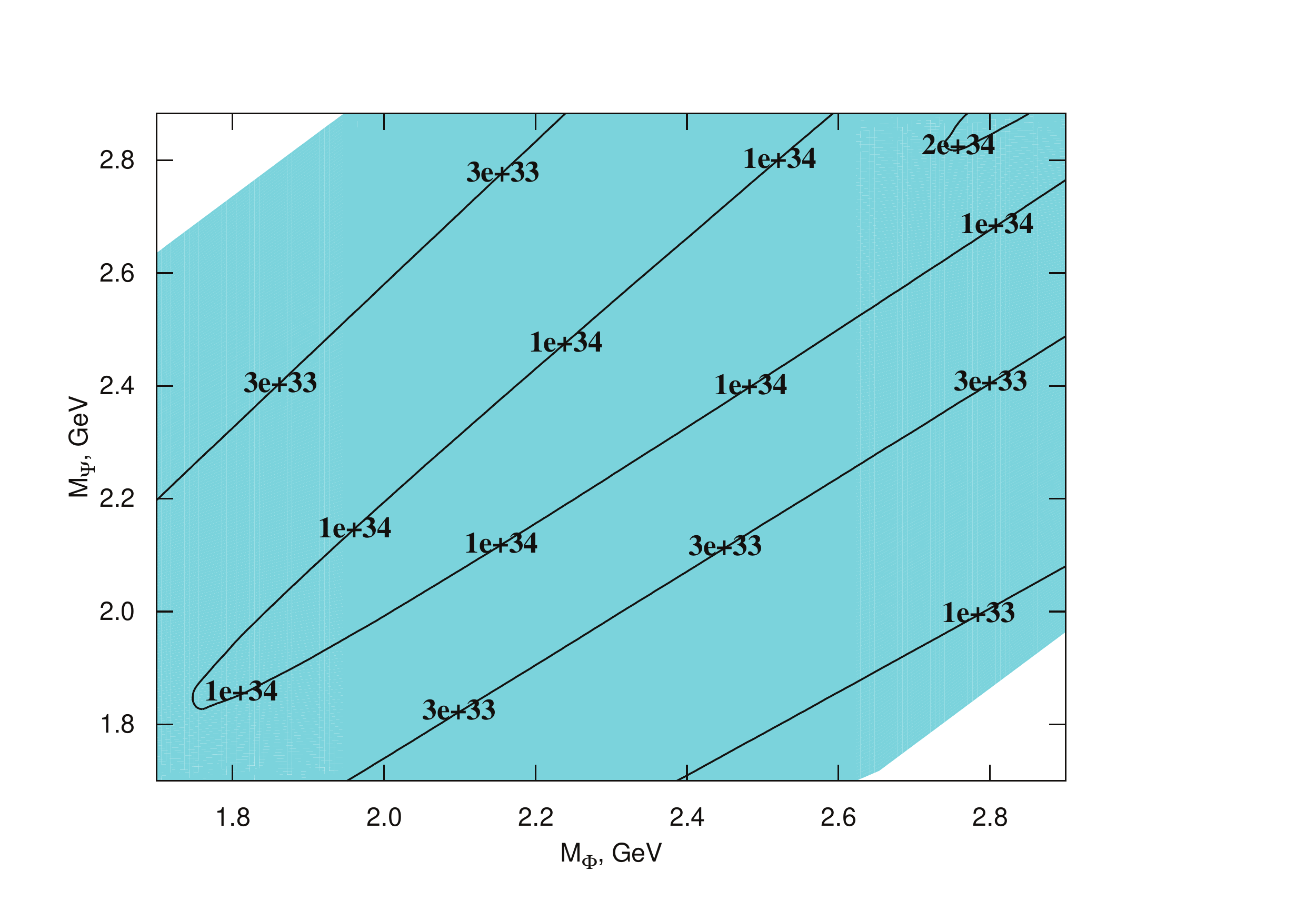} }
\caption{\label{fig:8} Contours of the constant lifetime (in years) of a
nucleon in the symmetric case with respect to IND with $\pi^0\eta$ in
the final state; we set $\Lambda=M_X=1$~TeV and $y=1$.} 
\end{figure}

For processes $\Phi+p\to \Psi+\pi^+\eta$ and
$\Phi+n\to \Psi+\pi^0\eta$, one multiplies the above expression for
the squared averaged matrix element by a factor of 2 and makes the
following substitutions
\begin{align*}
2\,p_{\pi^+}p_\Psi & =M \l M+2E_1-2E\r 
-M_\Psi^2-m_\pi^2+m_\eta^2\,,\\
p_Np_\Psi & =M_N (M-E)\,,\;\;\;
2\,p_\Psi p_{\eta} = M \l M-2E_1\r
 -M_\Psi^2+m_\pi^2-m_\eta^2\,,\\p_N p_{\pi^+}&= M_N E_1\,,\;\;
p_Np_\eta =\l E-E_1\r M_N\,,\;\;
2\,p_{\pi^+}p_{\eta}  =M \l 2E-M\r+
M_\Psi^2-m_\pi^2-m_\eta^2\,,
\end{align*}
and uses eqs.\,\eqref{con1}--\eqref{6.6} with 
\[
M=M_\Phi+M_N\,,\;\;\;m_1=m_\pi\,,\;\;\;m_2=m_\eta\,,\;\;\;m_3=M_\Psi\,.
\] 
Predictions for the proton lifetime for the $\pi^+\eta$ final state  
and neutron lifetime for $\pi^0\eta$ final state are presented for the
symmetric case in Fig,\,\ref{fig:7} and Fig.\,\ref{fig:8},
respectively. 
  
\begin{figure}[!htb]
\centerline{\includegraphics[width=0.8\columnwidth,angle=-90]{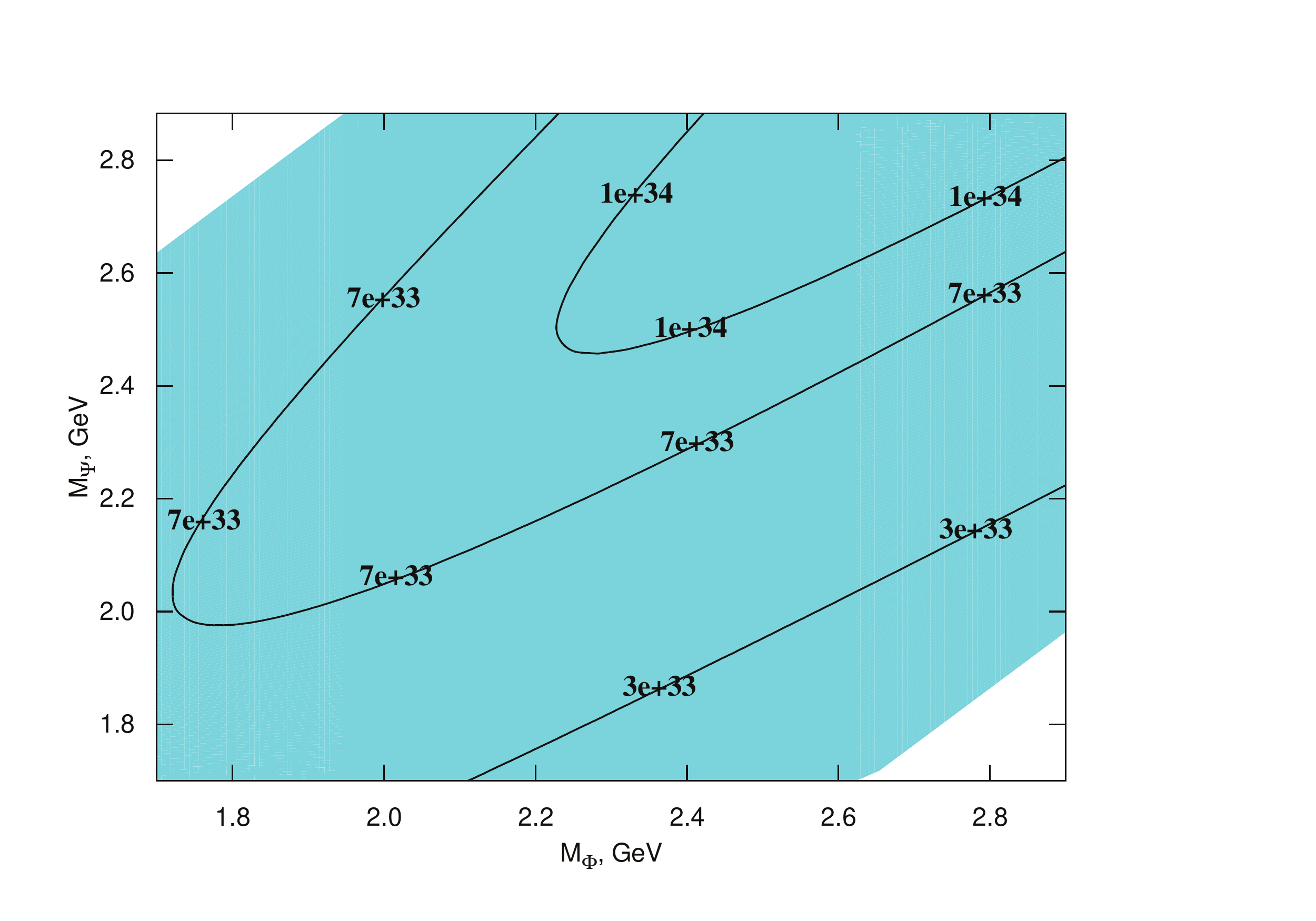} }
\caption{\label{fig:6} Contours of the constant lifetime (in years) of
a nucleon in the symmetric case with respect to IND process with
$\pi^0\pi^0$ in the final state; we set $\Lambda=M_X=1$~TeV and $y=1$.} 
\end{figure}

The squared matrix element for the processes $\Psi+n\to \Phi+2\eta$ and
$\Psi+n\to \Phi+2\pi^0$ averaged over spins of the initial particles
have the form  
\begin{equation}
\label{2neutrals}
\begin{split}
\overline{\left| {\cal M}\right|^2}=&2(A-B)^2\,p_Np_\Psi
+4B(A-B)\frac{M_N^2\,p_1p_\Psi}{I(p_N,p_1)}
+4B(A-B)\frac{M_N^2\,p_2p_\Psi}{I(p_N,p_2)}
\\&
+ 2B^2M_N^2\l
\frac{J(p_N,p_\Psi,p_1)}{I^2(p_N,p_1)}
+ \frac{J(p_N,p_\Psi,p_2)}{I^2(p_N,p_2)}
+ 2\frac{K(p_N,p_\Psi,p_1,p_2)}{I(p_N,p_1)I(p_N,p_2)}\r,
\end{split}
\end{equation}
where $p_1$ and $p_2$ are momenta of outgoing mesons and
\begin{equation}
A = \frac{3\beta c_1}{4f^2},\;\;\;
B = \frac{(3F-D)\beta c_1}{2f^2}
\end{equation}
for $\Psi+n\to \Phi+2\eta$ and 
\begin{equation}
A = \frac{\beta c_1}{4f^2},\;\;\;
B = \frac{(D+F)\beta c_1}{2f^2}
\end{equation}
for $\Psi+n\to \Phi+2\pi^0$.
In the laboratory frame one has 
\begin{equation}
\label{3d-decay-1a}
\begin{split}
&p_N p_1  = M_N E_1\,,\;\;
p_\Psi p_2=M_\Psi(E-E_1)\,,\;\;
2\,p_1p_2 =M \l 2E-M\r+
M_\Phi^2 - 2m_1^2\,,\\
&p_Np_2 =\l E-E_1\r M_N\,,\;\;\;p_Np_\Psi =M_N M_\Psi\,,\;\;\;
p_1p_\Psi =E_1 M_\Psi\,,
\end{split}
\end{equation}
and adopts eqs.\,\eqref{con1}--\eqref{6.6} with 
\begin{equation}
\label{3d-decay-1b}
M=M_\Psi+M_N\,, m_3=M_\Phi\,,\;\;\;\text{and}\;\;m_1=m_2=m_{\pi,\eta}\,.
\end{equation}
For the averaged squared matrix elements of 
$\Phi+n\to \Psi+2\eta$ and $\Phi+n\to \Psi+2\pi^0$, one has the same 
expression~\eqref{2neutrals} multiplied by a factor of two. In the
laboratory frame, one finds 
\begin{equation}
\label{3d-decay-2a}
\begin{split}
&2\,p_1p_\Psi =M \l M+2E_1-2E\r 
-M_\Psi^2\,,\;\;\;p_N p_1 = M_N E_1\,,\\ 
&2\,p_\Psi p_2 = M \l M-2E_1\r
 -M_\Psi^2\,,\;\;\;\,p_Np_\Psi =M_\Psi (M-E)\,,\\
&p_Np_2 =\l E-E_1\r M_N\,,\;\;\;
2\,p_1p_2 =M \l 2E-M\r+
M_\Psi^2-2m_{\pi,\eta}^2\,,
\end{split}
\end{equation}
and adopts eqs.\,\eqref{con1}--\eqref{6.6} with 
\begin{equation}
\label{3d-decay-2b}
M=M_\Phi+M_N\,, m_3=M_\Psi\,,\;\;\;\text{and}\;\;m_1=m_2=m_{\pi,\eta}\,.
\end{equation}
In Fig.\,\ref{fig:6} 
we present the predictions of neutron lifetime for the final state $\pi^0\pi^0$. 

\begin{figure}[!htb]
\centerline{\includegraphics[width=0.7\columnwidth,angle=-90]{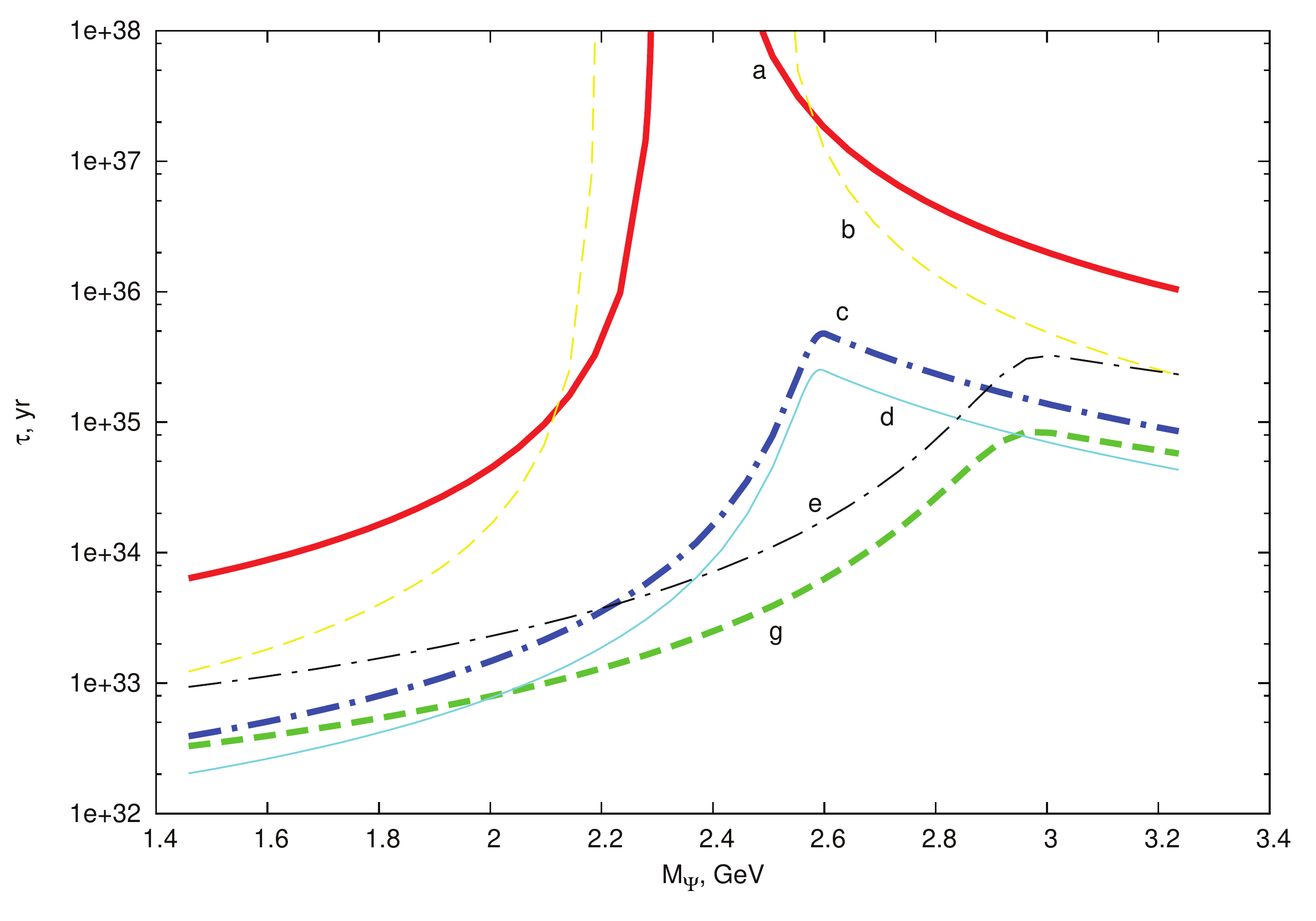} }
\caption{\label{fig:9} Contours of the constant lifetime (in years) of
the nucleon in the asymmetric case, 
$\eta=100$ ,  with respect to the two-meson processes: a)
$\bar{K}^0K^+$; b) $\eta\eta$; c) $\pi^0\eta$; d) $\pi^+\eta$; e)
$\pi^0\pi^0$; g) $\pi^+\pi^0$. Numbers for other processes with kaons
are similar to a) while for process with $\pi^+\pi^-$ are similar to
g). We set $\Lambda=M_X=1$~TeV and $y=1$. } 
\end{figure}

\begin{figure}[!htb]
\centerline{\includegraphics[width=0.7\columnwidth,angle=-90]{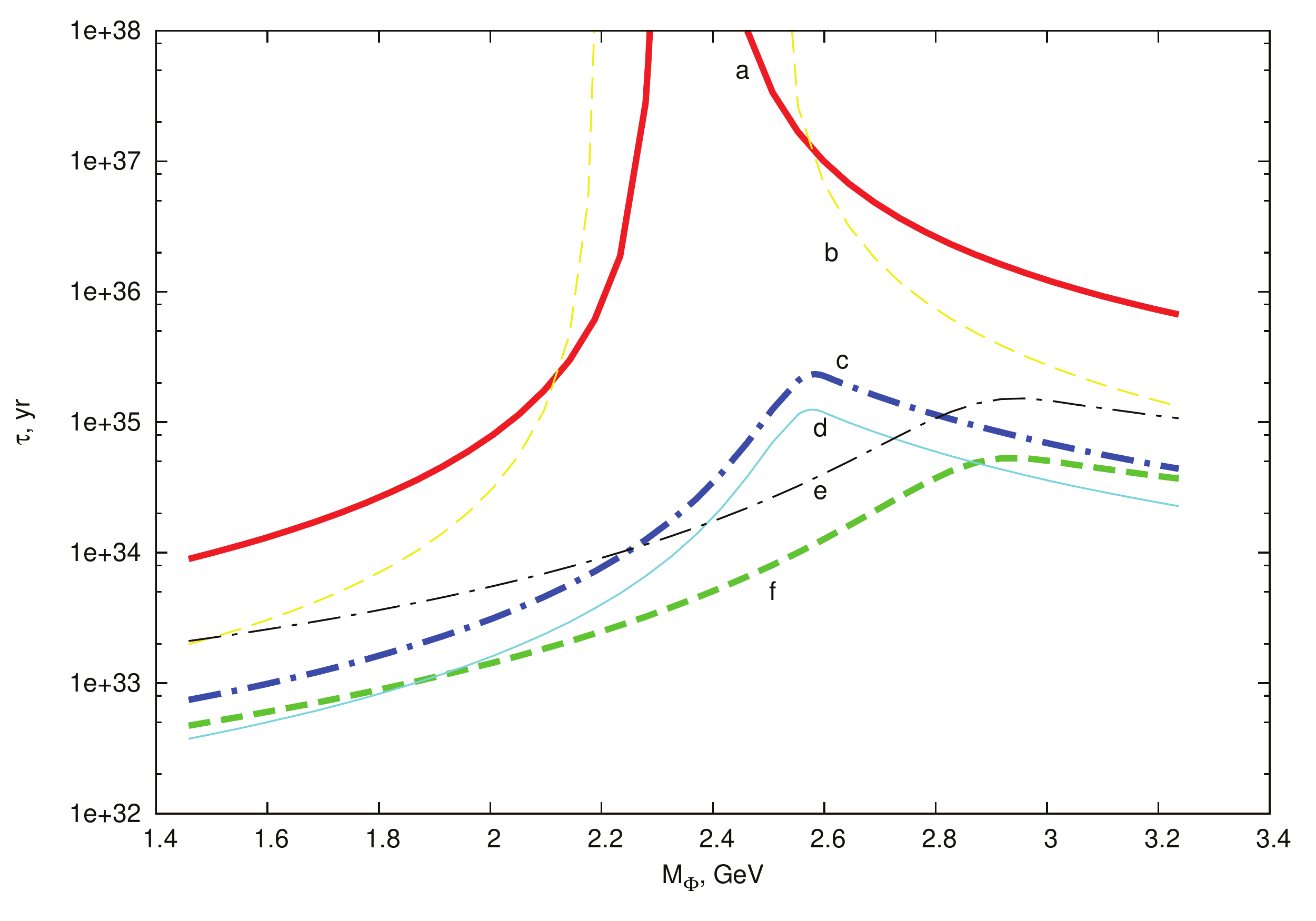}
} 
\caption{\label{fig:10} Contours of the constant lifetime (in years)
of a nucleon in the asymmetric 
case, $\eta=0.01$ ,  with respect to the two-meson processes. Other
notations are the same as in Fig.~\ref{fig:9}.} 
\end{figure}

Finally, to illustrate a dependence of the obtained predictions on the
value of nonspecified asymmetry between $\Psi$ and $\Phi$ populations,
$\eta$, we present in Figs.\,\ref{fig:9} and~\ref{fig:10}  
the estimates of the nucleon lifetime for two opposite cases of large
asymmetry $\eta=100$ and $\eta=0.01$, respectively. As one
observes, the predictions of nucleon lifetimes within hylogenesis
model can reach values around $10^{32}-10^{33}$~year which
looks quite promising for future experiments such as
Hyper-Kamiokande\,\cite{Abe:2011ts,HyperK} or DUNE\,\cite{DUNE}. 

The obtained predictions for double meson channels are, in general, 
only by an order of magnitude weaker than those for single-meson
channels (which can be as low as several units of
$10^{31}$~yr~\cite{Davoudiasl:2010am,Davoudiasl:2011fj} for the
same set of parameters). Note that the double meson signatures are
predicted for the proton decay in the context of grand unified
theories~\cite{Claudson:1981gh} as well as for dinucleon decays such 
as $pn\to \pi^+\pi^0$, for instance, in supersymmetric models with
R-parity violation; see e.g.,~\cite{Chemtob:2004xr}. Searches for the
latter type of processes had been performed by the Frejus
experiment~\cite{Berger:1991fa} and recently by the Super-Kamiokande
collaboration~\cite{Gustafson:2015qyo}. The most stringent limit for
the lifetime of the process $pn\to \pi^+\pi^0$ per oxygen nucleus is
found to be $\tau_{pn\to \pi^+\pi^0} > 1.70\times 10^{32}$~year. 
It has been obtained by making use of the expected kinematics of
dinucleon  decay. In particular, the angular distribution between
outgoing pions exhibits a maximum for events with back-to-back
topology and the distribution over momentum of $\pi^0$ has a
pronounced peak around nucleon 
mass. Because of this specific kinematics, the Super-Kamiokande result 
cannot be directly applied to the IND process with two pions in the final
state. However, one can show that for some combinations of masses of
dark matter particles,\footnote{In particular, when their
mass difference is large.} the distributions over momenta of outgoing
mesons also have maxima at $0.5-1$ GeV. This signature can be very
helpful in discriminating the IND process from the main background
which is the double pion production by atmospheric neutrinos. 
 

The double meson channels provide additional signatures of the
hylogenesis model, which will help to pin down the relevant model
parameters once the signal is found. Indeed, even the masses of dark
matter particles cannot be unambiguously extracted from a single-meson
event, because the initial nucleon momentum is not fixed in a real
experiment (the nucleon is not at rest); hence, the single mesons are
not monochromatic. A joint analysis of single and double meson events
can help to resolve the parameter values. Generally, one anticipates
that having more than one observable particle in the final state gives
more opportunities for background reduction in the future experiments.

The two-meson channels even can help to discriminate between proton
decay and induced proton decay, which may be challenging is some
situations. In particular, if single pions are registered at sub-GeV
range (say, below 500 MeV), an observation of multi-pion events with
higher total energies would favor the proton decay over the induced
proton decay in a model where the kinematics constrains the amount of
energy allocated to pion at sub-GeV range.

\section{Conclusions}
\label{Sec:conclusion}

Summarizing, in this paper we calculated the cross sections of several
IND processes for the hylogenesis model of dark matter. They include
the processes of mimicking neutron decays $n\to\nu\gamma$ and $n\to
e^+e^-$. Applying current best limits on the neutron lifetime with
respect to the processes $n\to\nu\gamma$ and $n\to e^+e^-$ and taking
into account the kinematics of the processes which were used in the
experiment, we obtained constraints on the parameter space of the
model. They are considerably weaker than the bounds obtained using the
results of the searches for events with a high $p_T$ jet and missing
energy signature at LHC experiments. 

Also, we calculated cross sections and lifetimes corresponding to IND
processes with two pions in the final state. Searches for such kinds
of signatures have not been performed yet and present an interesting
possibility to further explore the hylogenesis model. We found that
with the current bounds from the LHC data, the model allows for
a lifetime of IND such as $p\to \pi^+\pi^0$ or $n\to \pi^0\eta$ at the
level of $2\times 10^{32}$ yr. 

Note in passing, that by the time the new generation of
experiments looking for nucleon decay will be in operation, more data
from Run2 of the LHC allow an improvement of the collider sensitivity
to hylogenesis with respect to the analysis\,\cite{Demidov:2014mda}.

\vskip 0.3cm 
The work was supported by the RSCF grant 14-12-01430.

\appendix

\section{Couplings to baryons and mesons}
\label{App:A}
The interaction lagrangian of the type \eqref{contact} with the three
light quarks $q_1=u$, $q_2=d$, $q_3=s$
in terms of two-component spinors 
(the relevant are right-handed parts of the Dirac spinors) has 
the form\,\cite{Davoudiasl:2011fj} 
\begin{equation}
\label{appA:1}
{\cal L}_{int} = \text{Tr}\l {\cal C}\,{\cal O}\r+\text{h.c.}\,,\;\;\;\;
{\cal O}_{ij} \equiv \frac{1}{2}\,\Phi\,\epsilon_{\alpha\beta\gamma}\epsilon_{jkl}
\,\,{q_k}_{R}^{\alpha}{q_l}_R^{\beta}\,\,{q_i}_{R}^{\gamma}\Psi_R\,,
\end{equation}
where
\begin{equation}
\label{appA:2}
{\cal C} \equiv \l
\begin{array}{ccc}
\frac{c_2}{\sqrt{6}} + \frac{c_3}{\sqrt{2}} & 0 & 0\\
0 & \frac{c_2}{\sqrt{6}} - \frac{c_3}{\sqrt{2}} & 0\\
0 & c_1 & -\sqrt{\frac{2}{3}}\,c_2 \\
\end{array}
\r.
\end{equation}
The couplings $c_i$ are introduced as couplings to the three-quark
states which form the eigenstates of the strong isospin operator. 

Using the chiral perturbation theory one can obtain~\cite{Claudson:1981gh}
the corresponding interaction lagrangian for baryons 
\[
{\cal L}_{IND} = \beta\, \text{Tr} \l\Phi\xi {\cal C} \xi^{\dagger} \,
{\cal B}_R\Psi_R\r+\text{h.c.},
\]
where $\xi = {\rm exp}\left(i{\cal M}/f\right)$ and
\[
{\cal M} \equiv \left(
\begin{array}{ccc}
\frac{\eta}{\sqrt{6}}+\frac{\pi^0}{\sqrt{2}} & \pi^+ & K^{+}\\
\pi^- & \frac{\eta}{\sqrt{6}} - \frac{\pi^0}{\sqrt{2}} & K^0\\
K^- & \bar{K}^0 & -\sqrt{\frac{2}{3}}\eta \\
\end{array}
\right)
\]
and baryon fields leaving only a neutron and proton
\[
{\cal B}_R = \left(
\begin{array}{ccc}
0 & 0 & p_R\\
0 & 0 & n_R\\
0 & 0 & 0\\
\end{array}
\right).
\]
Expanding to linear order in meson fields we find (hereafter, in
terms of the Dirac fermions) 
\begin{equation}
\label{App:lagr0}
\begin{split}
{\cal L}_{1\pi} = \frac{i\beta}{f}\,\Phi\, \overline{\Psi^C} \,
\Bigg[&c_1\left(-\sqrt{\frac{3}{2}}n_R\,\eta + \frac{1}{\sqrt{2}}n_R\,\pi^0 -
p_R\,\pi^-\right)
\\
&+\left(\frac{c_2\sqrt{3}}{\sqrt{2}}
+ \frac{c_3}{\sqrt{2}}\right)p_R\,K^-
+\left(\frac{c_2\sqrt{3}}{\sqrt{2}} - \frac{c_3}{\sqrt{2}}\right)n_R\,\bar{K}^0
 \Bigg]+\text{h.c.}
\end{split}
\end{equation}
Expanding to the second order in $1/f$, one obtains 
\begin{equation}
\label{App:lagr1}
{\cal L}_{2\pi} = \frac{\beta}{2f^2}\l A_{31}\cdot \overline{\Psi^C}
p_R\,\Phi + A_{32}\cdot \overline{\Psi^C} n_R \,\Phi\r+\text{h.c.}
\end{equation}
where
\begin{align*}
A_{31}  = &-c_1\left(\sqrt{6}\pi^-\eta + K^0{K}^-\right)\\
&+\left(\frac{3}{2}c_2 + \frac{\sqrt{3}}{2}c_3\right)K^-\eta
+\left(\frac{\sqrt{3}}{2}c_2 + \frac{1}{2}c_3\right)K^-\pi^0
+\left(\sqrt{\frac{3}{2}}c_2 - \frac{3}{\sqrt{2}}c_3\right)\bar{K}^0\pi^-,\\
A_{32} = & -c_1\left(\pi^+\pi^- + \frac{3}{2}\eta^2
- \sqrt{3}\eta\pi^0 + \frac{1}{2}(\pi^{0})^2 + 2K^0\bar{K}^0
+ K^+K^-\right) \\ 
&+\left(\frac{\sqrt{3}}{\sqrt{2}}c_2 + \frac{3c_3}{\sqrt{2}}\right)K^-\pi^+
+\left(\frac{3}{2}c_2-\frac{\sqrt{3}}{2}c_3\right)\bar{K}^0\eta
-\left(\frac{\sqrt{3}}{2}c_2-\frac{c_3}{2}\right)\bar{K}^0\pi^0.
\end{align*}

Finally, for completeness let us remind~\cite{Claudson:1981gh} here the
interaction lagrangian of baryons with mesons to the leading order in
derivative expansion,  which has the form
\begin{equation}
\label{App:lagr_nucl}
\begin{split}
{\cal L} =
\frac{3F-D}{\sqrt{6}\,f}\l\bar{p}\gamma^{\mu}\gamma^5p + 
\bar{n}\gamma^{\mu}\gamma^5n\r\partial_{\mu}\eta 
+ \frac{D+F}{\sqrt{2}} 
\l\bar{p}\gamma^{\mu}\gamma^5p- \bar{n}\gamma^{\mu}\gamma^5n\r 
\partial_{\mu}\pi^0
\\
+\frac{D+F}{f}\l\partial_\mu\pi^+\bar{p}\gamma^{\mu}\gamma^5n
+\partial_\mu\pi^-\bar{n}\gamma^{\mu}\gamma^5p\r\;,
\end{split}
\end{equation}
where $D=0.8$ and $F=0.47$.
\bibliographystyle{plain}
\bibliography{ndecayJ_new}

\begin{thebibliography}{10}

\bibitem{Abe:2011ts}
K.~Abe et~al.
\newblock {Letter of Intent: The Hyper-Kamiokande Experiment --- Detector
  Design and Physics Potential ---}.
\newblock 2011.

\bibitem{Baer:2014eja}
Howard Baer, Ki-Young Choi, Jihn~E. Kim, and Leszek Roszkowski.
\newblock {Dark matter production in the early Universe: beyond the thermal
  WIMP paradigm}.
\newblock {\em Phys.Rept.}, 555:1--60, 2014.

\bibitem{Bell:2014xta}
Nicole~F. Bell, Shunsaku Horiuchi, and Ian~M. Shoemaker.
\newblock {Annihilating Asymmetric Dark Matter}.
\newblock {\em Phys.Rev.}, D91(2):023505, 2015.

\bibitem{Berger:1991fa}
Christoph Berger et~al.
\newblock {Lifetime limits on (B-L) violating nucleon decay and dinucleon decay
  modes from the Frejus experiment}.
\newblock {\em Phys. Lett.}, B269:227--233, 1991.

\bibitem{Blewitt:1985zg}
G.~Blewitt, H.S. Park, B.G. Cortez, G.W. Foster, W.~Gajewski, et~al.
\newblock {Experimental Limits on the Nucleon Lifetime for Two and Three-body
  Decay Modes}.
\newblock {\em Phys.Rev.Lett.}, 54:22, 1985.

\bibitem{Blinov:2012hq}
Nikita Blinov, David~E. Morrissey, Kris Sigurdson, and Sean Tulin.
\newblock {Dark Matter Antibaryons from a Supersymmetric Hidden Sector}.
\newblock {\em Phys.Rev.}, D86:095021, 2012.

\bibitem{Chemtob:2004xr}
Marc Chemtob.
\newblock {Phenomenological constraints on broken R parity symmetry in
  supersymmetry models}.
\newblock {\em Prog. Part. Nucl. Phys.}, 54:71--191, 2005.

\bibitem{Claudson:1981gh}
Mark Claudson, Mark~B. Wise, and Lawrence~J. Hall.
\newblock {Chiral Lagrangian for Deep Mine Physics}.
\newblock {\em Nucl.Phys.}, B195:297, 1982.

\bibitem{Davoudiasl:2010am}
Hooman Davoudiasl, David~E. Morrissey, Kris Sigurdson, and Sean Tulin.
\newblock {Hylogenesis: A Unified Origin for Baryonic Visible Matter and
  Antibaryonic Dark Matter}.
\newblock {\em Phys.Rev.Lett.}, 105:211304, 2010.

\bibitem{Davoudiasl:2011fj}
Hooman Davoudiasl, David~E. Morrissey, Kris Sigurdson, and Sean Tulin.
\newblock {Baryon Destruction by Asymmetric Dark Matter}.
\newblock {\em Phys.Rev.}, D84:096008, 2011.

\bibitem{Demidov:2014mda}
S.V. Demidov, D.S. Gorbunov, and D.V. Kirpichnikov.
\newblock {Collider signatures of Hylogenesis}.
\newblock {\em Phys.Rev.}, D91(3):035005, 2015.

\bibitem{Fukugita:1986hr}
M.~Fukugita and T.~Yanagida.
\newblock {Baryogenesis Without Grand Unification}.
\newblock {\em Phys.Lett.}, B174:45, 1986.

\bibitem{Gorbunov:2012ij}
D.S. Gorbunov and A.G. Panin.
\newblock {Free scalar dark matter candidates in $R^2$-inflation: the light,
  the heavy and the superheavy}.
\newblock {\em Phys.Lett.}, B718:15--20, 2012.

\bibitem{Gustafson:2015qyo}
J.~Gustafson et~al.
\newblock {Search for dinucleon decay into pions at Super-Kamiokande}.
\newblock {\em Phys. Rev.}, D91(7):072009, 2015.

\bibitem{Hardy:2014dea}
Edward Hardy, Robert Lasenby, and James Unwin.
\newblock {Annihilation Signals from Asymmetric Dark Matter}.
\newblock {\em JHEP}, 07:049, 2014.

\bibitem{HyperK}
http://http://www.hyper k.org/en/.
\newblock {}.
\newblock 2011.

\bibitem{DUNE}
http://www.dunescience.org/.
\newblock {}.
\newblock 2015.

\bibitem{Hu:2000ke}
Wayne Hu, Rennan Barkana, and Andrei Gruzinov.
\newblock {Cold and fuzzy dark matter}.
\newblock {\em Phys.Rev.Lett.}, 85:1158--1161, 2000.

\bibitem{McGrew:1999nd}
C.~McGrew, R.~Becker-Szendy, C.B. Bratton, J.L. Breault, D.R. Cady, et~al.
\newblock {Search for nucleon decay using the IMB-3 detector}.
\newblock {\em Phys.Rev.}, D59:052004, 1999.

\bibitem{Agashe:2014kda}
K.A. Olive et~al.
\newblock {Review of Particle Physics}.
\newblock {\em Chin.Phys.}, C38:090001, 2014.

\bibitem{Petraki:2013wwa}
Kalliopi Petraki and Raymond~R. Volkas.
\newblock {Review of asymmetric dark matter}.
\newblock {\em Int.J.Mod.Phys.}, A28:1330028, 2013.

\end{thebibliography}

\end{document}